\documentclass[final,reqno]{amsart}

\usepackage{amsaddr}
\usepackage{amsmath,amssymb,amsfonts,amsthm}
\usepackage{enumitem}
\usepackage{xcolor}
\usepackage{tikz}
\usepackage{graphicx}
\usepackage{hyperref}

\hypersetup{hidelinks}

\newtheorem{theorem}{Theorem}[section]

\newtheorem{corollary}[theorem]{Corollary}
\newtheorem{proposition}[theorem]{Proposition}
\theoremstyle{remark}
\newtheorem{remark}[theorem]{Remark}
\newtheorem*{remark*}{Remark}

\DeclareMathOperator{\Span}{Span}
\DeclareMathOperator{\cl}{cl}
\DeclareMathOperator{\dom}{dom}
\DeclareMathOperator{\inte}{int}
\DeclareMathOperator{\Tr}{Tr}

\newcommand{\RR}{\mathbb{R}}
\newcommand{\NN}{\mathcal{N}}
\newcommand{\NC}{n}
\newcommand{\NF}{\mathfrak{N}}
\newcommand{\CF}{\mathfrak{C}}
\newcommand{\dua}[2]{\langle #1,#2\rangle}
\newcommand{\norm}[1]{\left\lVert #1\right\rVert}
\newcommand{\vertiii}[1]{\left\vert\kern-0.25ex\left\vert\kern-0.25ex
  \left\vert #1\right\vert\kern-0.25ex\right\vert\kern-0.25ex\right\vert}
\newcommand{\dd}{\mathrm{d}}

\renewcommand{\epsilon}{\varepsilon}
\renewcommand{\phi}{\varphi}

\renewcommand{\le}{\leqslant}
\renewcommand{\ge}{\geqslant}

\title[Neural network approximation in DFT]
{Neural network approximation of regularized density functionals}

\author{ Andre Laestadius$^{1,2,3}$
\and Mathias Oster$^4$}

\address{$^1$Department of Mathematical Sciences, Chalmers University of
Technology and the University of Gothenburg, G\"oteborg, Sweden}

\address{$^2$Hylleraas Centre for Quantum Molecular Sciences, Department of
Chemistry, University of Oslo, Oslo, Norway}

\address{$^3$Department of Computer Science, Oslo Metropolitan University, Oslo, Norway}

\address{$^4$Institut f\"ur Geometrie und Praktische Mathematik, RWTH Aachen
University, Aachen, Germany}

\begin{document}

\begin{abstract}
We study neural-network approximation of density functionals
in a Banach-space setting.  Since the exact continuum functionals are
discontinuous in the natural density topology, we apply $p$-Moreau--Yosida regularization. We construct finite-rank
ReLU networks that preserve convexity and nonnegativity, lie below the
regularized functional, and approximate it uniformly on compact sets.
For energies obtained by minimization over compact density sets, we account
exactly for the energy shift caused by regularization and bound the remaining
error by the localization on these compact sets and network approximation errors. We apply this framework
to canonical and grand-canonical density-functional
theory at fixed particle number and fixed mean particle number,
respectively. We obtain uniform energy
convergence on norm-compact classes of external potentials, including
molecular Coulomb potentials.  We illustrate our results through
numerical experiments on a simple model system.
\end{abstract}

\maketitle

\section{Introduction}\label{sec:introduction}

Density-functional theory (DFT) expresses the ground-state energy of a
quantum many-electron system through a variational principle involving only
the one-body electron density.  Central to this formulation is the universal
density functional, which contains the kinetic and interaction energies
and is independent of the external potential
\cite{Hohenberg1964,lieb1983density}.
The exact universal functional is convex, with the ground-state energy
obtained by minimizing the sum of this functional and the linear external-potential term.
Nevertheless, the exact evaluation of the universal functional requires
solving a many-body problem, and the construction of useful approximations
remains a central task in DFT
\cite{burke2012perspective,teale2022dft}.

Machine learning has opened another route to such approximations, from
learning kinetic-energy functionals \cite{snyder2012finding} to
neural-network exchange--correlation functionals
\cite{nagai2020completing,kirkpatrick2021pushing}. A review of computational
approaches is given in \cite{fiedler2022deep}.
These approaches also pose mathematical challenges. First, agreement with
the exact functional at sampled densities gives
no general knowledge of the approximation errors on other data points.
A general network need not retain the convexity of the
universal functional.
Moreover, the exact functional is discontinuous in the ``natural''
$L^1\cap L^3$ topology \cite{lammert2007differentiability}, so that universal
approximation results for continuous functionals cannot be applied
directly.

The principal aim of this paper is to establish a mathematical framework
for neural-network approximation of regularized density functionals and
their associated ground-state energies. Following the convex-analytic
formulation of DFT \cite{lieb1983density}, we use $p$-Moreau--Yosida
regularization \cite{kvaal2014differentiable,MYperspective} to obtain
continuous convex functionals. We then construct convex networks and
control the energies obtained by minimizing the regularized functionals
over compact sets of densities.

A novelty of this work is that the regularization is performed with the
particle-number constraint in place, in contrast to
\cite{kvaal2014differentiable}, where the Moreau--Yosida envelope is taken
over the whole space.  As a result, the effect of the regularization on the
energy depends only on the external potential modulo an additive constant.
In this way, the regularized energy has the same gauge covariance as the
exact energy, i.e., shifting the external potential by a constant $c$ shifts
both energies by $cN$.

\subsection*{Contributions}
\noindent Our main results can be summarized as follows.
\begin{enumerate}[label=\textup{(\roman*)}]
 \item \emph{Convex finite-rank approximation.}
 Every continuous nonnegative convex functional on a real normed space admits
 convex finite-rank ReLU networks that are nonnegative global lower bounds and
 approximate it uniformly on any prescribed compact set
 (Theorem~\ref{thm:UAPconvex}).
 \item \emph{Restricted regularization on nonreflexive spaces.}
 We study $p$-Moreau--Yosida regularization over closed convex sets,
 with particular attention to affine constraints fixing the particle number.
 Differentiability of the regularized functional is obtained
 through duality from strict convexity of the dual norm, rather than
 through proximal attainment, which may fail on nonreflexive spaces
 (Theorem~\ref{thm:MYBacho}). For affine
 constraints, we identify the exact regularization shift in terms of the quotient dual norm
 (Proposition~\ref{prop:exact-affine-shift}).
 \item \emph{Energy error estimates.}
 For energies defined through a concave conjugate, we compare the exact
 energy with the network energy, obtained by minimizing a convex network
 approximation of the regularized functional over a compact set of
 densities.
 After correcting the exact regularization shift, we bound the energy error
 by the localization error on compact sets and the uniform network
 approximation error (Theorem~\ref{thm:ErrorLegendre}).
 \item \emph{Application to DFT.}
 We apply the construction to canonical and grand-canonical density
 functionals at fixed particle number.  By choosing the localization sets according to the desired
 accuracy, we obtain the uniform convergence of the energy on norm-compact classes
 of external potentials, including molecular Coulomb potentials
 (Proposition~\ref{prop:dft-compact-potential-classes}).
\end{enumerate}

Section~\ref{sec:abstract-framework} develops the abstract framework, and
Section~\ref{sec:dft-application} specializes it to DFT.
Section~\ref{sec:numerical-experiments} illustrates the construction for
two fermions on a four-site ring.  Proofs are collected in
Section~\ref{proofsec}.

\subsection*{Acknowledgments}

\noindent The first author acknowledges funding from the European Research Council
through ERC-2021-STG Grant Agreement No.~101041487 (REGAL) as well as funding from the Research Council of
Norway (RCN) through the Hylleraas Centre for Quantum Molecular Sciences
(Centre of Excellence, Grant No.~262695), and through project No.~361617
(QSTAR, Centre for Quantum Computing and Applications).

The authors acknowledge the fruitful discussions with Mih\'aly A. Csirik.

\subsection*{AI use disclosure}

\noindent During the preparation of this work the authors made use of
the large language models Claude (Anthropic) and ChatGPT (OpenAI).  Beyond
editing and restructuring the text, these models were used to check mathematical
arguments, and to identify gaps and errors. Additionally, they were used to design a concrete counterexample in the spirit of Lammert's work \cite{lammert2007differentiability}.  All output was reviewed,
verified and edited by the authors, who take full responsibility for the
content of this publication.

\section{Abstract approximation framework}\label{sec:abstract-framework}

\subsection{Neural-network approximation}\label{absnn}

We first work with a real separable Banach space $X$ and recall the
Banach-space universal approximation theorem of
\cite{benth2021neural}.  For a continuous map $\sigma:X\to X$, called an
\emph{activation function}, $\ell\in X^*$, $A\in\mathcal L(X)$, and $b\in X$,
define the \emph{neuron}
\[
 \NC_{\ell,A,b}(\rho)=\dua{\ell}{\sigma(A\rho+b)},\qquad \rho\in X,
\]
and the associated class of neural networks
\[
 \NF(\sigma)=\Span\{\NC_{\ell,A,b}:\ell\in X^*,\
                  A\in\mathcal L(X),\ b\in X\}.
\]
Throughout the manuscript, all sets denoted by $K$, including indexed
variants, are assumed to be nonempty. The activation function $\sigma$ is called \emph{discriminatory} if, for
every compact $K\subset X$ and every finite signed regular Borel measure
$\mu$ on $K$,
\[
 \int_K\dua{\ell}{\sigma(A\rho+b)}\,\mu(\dd\rho)=0
 \quad\text{for all }\ell\in X^*,\ A\in\mathcal L(X),\ b\in X
 \quad\Longrightarrow\quad \mu=0.
\]

We equip $C(X,\RR)$ with the compact-open topology generated by the
seminorms
\[
 |G|_K=\sup_{\rho\in K}|G(\rho)|,
 \qquad K\subset X\ \text{compact}.
\]
All closures and interiors of subsets of $C(X,\RR)$ in this subsection are
taken in this topology.

\begin{theorem}[Universal approximation property, \cite{benth2021neural}]\label{uap}
Suppose that $\sigma:X\to X$ is continuous and discriminatory.  Then
$\NF(\sigma)$ is dense in $C(X,\RR)$ for the compact-open topology.  More
explicitly, for every $G\in C(X,\RR)$, compact $K\subset X$, and
$\delta>0$, there exists $\NN\in\NF(\sigma)$ such that
\[
 |G-\NN|_K<\delta.
\]
\end{theorem}
\begin{remark}
If $X$ admits a normalized Schauder basis and $\sigma$ is Lipschitz
continuous, the construction of
\cite[Proposition~4.1]{benth2021neural} approximates the network in
Theorem~\ref{uap} through finite-dimensional coordinate projections.
\end{remark}

Note that the network
 depends on $K$, and Theorem~\ref{uap} provides no control away from
$K$. Furthermore, uniform approximation alone does not ensure that an approximating network
inherits the target functional's additional structure like nonnegativity or convexity.
The following result gives a general way to retain a constraint when
the corresponding class of admissible functionals has nonempty interior.

\begin{theorem}[Constrained UAP]\label{sigapprox}
Suppose that $\sigma:X\to X$ is continuous and discriminatory, and let
$\CF\subset C(X,\RR)$ satisfy
\[
 \CF\subset\cl(\inte\CF).
\]
Then $\NF(\sigma)\cap\inte\CF$ is dense in $\CF$.  Thus, for every
$G\in\CF$, compact $K\subset X$, and $\delta>0$, there exists
\[
 \NN\in\NF(\sigma)\cap\inte\CF
 \qquad\text{such that}\qquad |G-\NN|_K<\delta.
\]
\end{theorem}

Nonnegativity on a fixed compact set meets this hypothesis.
The next proposition makes this observation precise.

\begin{proposition}
\label{prop:Positivity}
For a compact $K\subset X$, let
\[
 \CF_K^+=\{G\in C(X,\RR):G(\rho)\ge0\text{ for every }\rho\in K\}.
\]
Then $\CF_K^+$ is closed and
\[
 \inte\CF_K^+=\{G\in C(X,\RR):\min_KG>0\},
 \qquad
 \CF_K^+=\cl(\inte\CF_K^+).
\]
\end{proposition}

Convexity is another property that we want the approximating functional
to retain.  On a Banach space, the class of globally
convex continuous functionals has empty interior in the compact-open
topology, i.e., perturbations away from a prescribed compact set can destroy
convexity.  Thus  Theorem~\ref{sigapprox}
does not apply to this constraint.

We now work on an arbitrary real normed space $X$.
We build convexity directly into the approximating functional by taking it to be the maximum of
finitely many affine functions defined by a set of supporting hyperplanes of the
target functional.  Such a maximum is convex on all of $X$ and can be
evaluated using the scalar activation
$\operatorname{ReLU}(t)=\max\{0,t\}$ through the identity
\[
 \max\{s,t\}=\operatorname{ReLU}(s-t)+t.
\]
A map obtained
by first applying a finite-dimensional affine feature map
\[
 \rho\longmapsto
 \bigl(\dua{a_1}{\rho}+b_1,\ldots,\dua{a_M}{\rho}+b_M\bigr)\in\RR^M
\]
and then an ordinary finite-dimensional ReLU network will be called a \emph{finite-rank ReLU network} on $X$.

\begin{theorem}[Convex finite-rank approximation]
\label{thm:UAPconvex}
Let $X$ be a real normed space and let $G:X\to[0,+\infty)$ be continuous and
convex.  For every compact $K\subset X$ and $\delta>0$, there are $M\in\mathbb N$,
$a_i\in X^*$ and $b_i\in\RR$, $i=1,\ldots,M$, such that
\begin{equation}\label{eq:convex-network}
 \NN(\rho)=\max\{0,\dua{a_1}{\rho}+b_1,\ldots,
                    \dua{a_M}{\rho}+b_M\}
\end{equation}
is a finite-rank ReLU network satisfying
\[
 0\le\NN(\rho)\le G(\rho)\quad(\rho\in X),
 \qquad
 0\le G(\rho)-\NN(\rho)<\delta\quad(\rho\in K).
\]
\end{theorem}

\begin{remark}
Under the hypotheses of Theorem~\ref{thm:UAPconvex}, let
$(\psi_\lambda)_{\lambda\in\Lambda}$ be a family of continuous convex functions
$X\to\RR$ satisfying
\begin{equation}\label{eq:supGcond}
 \psi_\lambda\le G\quad\text{on }X,
 \qquad
 G(\rho)=\sup_{\lambda\in\Lambda}\psi_\lambda(\rho)
 \quad(\rho\in K).
\end{equation}
 Select features $\psi_1,\ldots,\psi_M$ and define
\[
 \NN(\rho)=\max\{0,\psi_1(\rho),\ldots,\psi_M(\rho)\}.
\]
Then $\NN$ is continuous, convex, and satisfies the same approximation bounds.
Thus the affine feature map can be replaced by a non-affine one. However,
only the existence of an affine family satisfying \eqref{eq:supGcond} is guaranteed.
The choice $\psi_1=G$ gives a trivial exact representation. Approximation
using a prescribed, more general family of features is not treated here.
\end{remark}

Theorem~\ref{thm:UAPconvex} gives a nonnegative convex lower approximant of
$G$ on all of $X$ that approximates $G$ uniformly on $K$.  The proof combines
supporting affine functions with a finite subcover of $K$.
It requires neither completeness nor separability of $X$, and no
discriminatory activation is assumed.

\begin{remark}
\label{rem:network-localization}
The convex finite-rank approximations with affine feature maps $\langle a_i,\rho\rangle$ (or Lipschitz pieces $\psi_i$) are globally Lipschitz. Therefore, the domain of each Fenchel conjugate is bounded and, thus, there are $v\in X^*$ such that $$\inf_{\rho\in X}\{\NN(\rho)+\dua{v}{\rho}\}=-\infty.$$ To be precise, convex programming gives
\begin{equation*}
 \inf_{\rho\in X}\{\NN(\rho)+\dua{v}{\rho}\}>-\infty
 \quad\Longleftrightarrow\quad -v\in C,
 \qquad v\in X^*,
\end{equation*} where $C=\operatorname{conv}\{0,a_1,\ldots,a_M\}\subset X^*$.
When $X$ is infinite dimensional, the finite-dimensional set $C$ has empty
interior in $X^*$.  A fixed network of this form therefore cannot give
finite unrestricted energies on any ball of potentials.
For networks defined on the tangent space $Y$ of an affine slice
$\bar\rho+Y$, the same criterion uses the restricted potential $v|_Y$.
The constant $\dua{v}{\bar\rho}$ does not affect finiteness.
This explains the compact localization used in
Section~\ref{sec:abstract-error}: on a compact set $K$, the
continuous network objective has a finite attained minimum.  If $K$ is also
convex, its minimization is a convex optimization problem.
Figure~\ref{fig:convex-lower-network} illustrates the failure of
unrestricted minimization even for a one-dimensional convex lower
approximant.
\end{remark}

\begin{figure}[t]
\centering
\begin{tikzpicture}[x=0.61cm,y=0.20cm,>=stealth,font=\footnotesize]
  \path[use as bounding box] (-5.6,-10) rectangle (3,13.4);
  \fill[gray!8] (-2.4,-8.2) rectangle (2.4,11.5);
  \draw[gray!55,densely dotted] (-2.4,-8.8) -- (-2.4,11.5);
  \draw[gray!55,densely dotted] (2.4,-8.8) -- (2.4,11.5);
  \draw[very thick,gray!75] (-2.4,-8.8) -- (2.4,-8.8);
  \draw[gray!75] (-2.4,-9.1) -- (-2.4,-8.5);
  \draw[gray!75] (2.4,-9.1) -- (2.4,-8.5);
  \node[below] at (0,-8.8) {$K$};
  \draw[->,black] (-5.35,0) -- (2.8,0) node[right] {$\rho$};
  \draw[->,black] (0,-8.2) -- (0,11.8);
  \draw[black] (-3,-0.3) -- (-3,0.3) node[above] {$-3$};
  \node[above] at (-1.3,11.9) {$v=3,\quad -v\notin[-2,2]$};

  \draw[blue!70!black,very thick,domain=-5.2:2.65,samples=100,smooth]
    plot (\x,{0.5*\x*\x+3*\x});
  \node[blue!70!black] at (-4.05,4.1) {$G+3\rho$};

  \draw[orange!85!black,very thick]
    (-5.2,-7.2) -- (-1,-3) -- (1,3) -- (2.65,11.25);
  \draw[->,orange!85!black,very thick] (-4.25,-6.25) -- (-5.2,-7.2);
  \node[orange!85!black,below] at (-4.65,-7.25) {$-\infty$};
  \node[orange!85!black] at (1.25,-3.5) {$\NN+3\rho$};

  \fill[black] (-2.4,-4.4) circle (1.3pt);
  \draw[->,black] (-0.5,-6.5) -- (-2.25,-4.6);
  \node[right] at (-0.5,-6.5) {minimum on $K$};
\end{tikzpicture}
\caption{Illustration of failure for unrestricted minimization and the effect of compact
 localization.
}
\label{fig:convex-lower-network}
\end{figure}

\begin{remark}
The construction in Theorem~\ref{thm:UAPconvex} uses finitely many supporting
hyperplanes selected by compactness.  Bounds on the number $M$
of supporting affine functions in terms of
$\delta$ require additional information: if the subgradients of $G$ are
bounded in norm by $L$ on a neighborhood of $K$, then, for all sufficiently
small $\delta$, one can take $M\le\operatorname{cov}(K,\delta/(2L))$, the minimal number of
open balls of radius $\delta/(2L)$ covering $K$.  In finite dimensions, the same
argument applies to a nonnegative proper lower-semicontinuous convex
functional $G:X\to[0,+\infty]$ when $K\subset\inte(\dom G)$.
\end{remark}

\subsection{\texorpdfstring{$p$-Moreau--Yosida regularization}{p-Moreau-Yosida regularization}}
\label{mysec}

Fix a real normed space $X$ equipped with a norm
$\norm{\cdot}_X$ and dual norm $\norm{\cdot}_{X^*}$.
Throughout the manuscript, unless otherwise stated, we fix $1<p<\infty$
and let $q=p/(p-1)$ denote its H\"older conjugate.
Throughout the theoretical development and the proofs, dependence on this
fixed $p$ is suppressed in the notation.
For $\epsilon>0$, set
\[
 \phi_{\epsilon}(\rho)=\frac{1}{p\epsilon}\norm{\rho}_X^p.
\]
For a proper convex functional $G:X\to[0,+\infty]$, its \emph{$p$-Moreau--Yosida
($p$-MY) regularization} is
\begin{equation}\label{eq:abstract-envelope}
 G^{\epsilon}(\rho)
 =\inf_{\sigma\in X}
 \left\{G(\sigma)+\frac{1}{p\epsilon}
                   \norm{\rho-\sigma}_X^p\right\}
 =G\mathbin{\square}\phi_{\epsilon}(\rho).
\end{equation}
For an extended-real functional $\Phi:X\to(-\infty,+\infty]$, its
\emph{convex conjugate} is
\[
 \Phi^*(v)=\sup_{\rho\in X}\bigl\{\dua{v}{\rho}-\Phi(\rho)\bigr\},
 \qquad v\in X^*.
\]
The scaling of the penalty term is chosen so that
\begin{equation}\label{eq:power-conjugate}
 \phi_{\epsilon}^*(v)
 =\frac{\epsilon^{q-1}}{q}\norm{v}_{X^*}^q,
\end{equation}
and the quadratic regularization is recovered when $p=q=2$.
Bacho \cite{Bacho2023} studies $p$-MY envelopes, including basic regularity
on normed spaces and differentiability on reflexive Banach spaces.
For $p=2$, Khanh and Nguyen \cite{KhanhNguyen2022} prove
G\^ateaux differentiability for proper lower-semicontinuous convex
functionals on Banach spaces with strictly convex dual norm, without
assuming reflexivity.  The formulation below collects the basic properties
on normed spaces for general $p>1$.
\begin{theorem}
\label{thm:MYBacho}
Let $X$ be a real normed space and let $G:X\to[0,+\infty]$ be proper and convex.
\begin{enumerate}[label=\textup{(\roman*)}]
 \item $G^{\epsilon}$ is finite, convex, and locally Lipschitz continuous
       on $X$.  Consequently,
       $\partial G^{\epsilon}(\rho)\ne\varnothing$ for every $\rho\in X$.
 \item If $0<\delta<\epsilon$, then, pointwise on $X$,
 \[
  \inf_XG\le G^{\epsilon}\le G^{\delta}\le G,
  \qquad
  \inf_XG^{\epsilon}=\inf_XG.
 \]
 \item If $G$ is lower semicontinuous in the norm topology, then
       $G^{\epsilon}(\rho)\nearrow G(\rho)$ as $\epsilon\downarrow0$ for
       every $\rho\in X$.
\item If $\norm{\cdot}_{X^*}$ is strictly convex, then
       $G^{\epsilon}$ is G\^ateaux differentiable on $X$.
\end{enumerate}
\end{theorem}

The differentiability conclusion in (iv) does not require the infimum in
\eqref{eq:abstract-envelope} to be attained.  Instead, it follows from the identity
\begin{equation*}
 (G^{\epsilon})^*(v)
 =G^*(v)+\frac{\epsilon^{q-1}}{q}\norm{v}_{X^*}^q
\end{equation*}
by Fenchel reciprocity.

For the remainder of this section, assume that $X$ is a real Banach space.
On separable spaces, the strict convexity assumed in
Theorem~\ref{thm:MYBacho}(iv) can always be achieved according to the following
classical result, see \cite[Theorem~8.13]{fabian2001functional}.

\begin{proposition}[Theorem~8.13 in~\cite{fabian2001functional}]
\label{prop:dual-renorming}
Every separable real Banach space admits an equivalent norm whose dual norm is
strictly convex.  The corresponding primal norm is G\^ateaux differentiable
away from the origin.
\end{proposition}

In particular, differentiability in Theorem~\ref{thm:MYBacho} can always be obtained
after an equivalent renorming when $X$ is separable.

We next record the restricted and translated forms of the regularization.  Let
$G:X\to[0,+\infty]$ be proper and convex, and let
$\mathcal A\subset X$ be nonempty, closed and convex with
$\mathcal A\cap\dom G\ne\varnothing$.  Let $\iota_{\mathcal A}$ be the
indicator function equal to zero on $\mathcal A$ and $+\infty$ outside
$\mathcal A$.  The restricted regularization is
\begin{equation}\label{eq:restricted-envelope}
 (G+\iota_{\mathcal A})^{\epsilon}(\rho)
 =\inf_{\sigma\in\mathcal A}
 \left\{G(\sigma)+\frac{1}{p\epsilon}
                   \norm{\rho-\sigma}_X^p\right\}.
\end{equation}
The indicator is added before regularizing because, in the applications of
Section~\ref{sec:dft-application}, the functional is minimized only over a
constraint set $\mathcal A$, such as a fixed-mass slice.  The infimum in
\eqref{eq:restricted-envelope} runs over $\sigma\in\mathcal A$ only, so the
regularized functional depends only on the values of $G$ on $\mathcal A$.
If $G$ is finite outside $\mathcal A$, it differs in general from the
restriction of $G^{\epsilon}$ to $\mathcal A$.

\begin{proposition}
\label{prop:restricted-envelope}
The following properties hold true.
\begin{enumerate}[label=\textup{(\alph*)}]
 \item The conclusions of Theorem~\ref{thm:MYBacho}(i) and (ii) apply to
       $(G+\iota_{\mathcal A})^{\epsilon}$.
 \item If $G+\iota_{\mathcal A}$ is lower semicontinuous in the norm
       topology, then (iii) applies.
 \item If $\norm{\cdot}_{X^*}$ is strictly convex, then conclusion~(iv)
       applies.
\end{enumerate}
\end{proposition}

When
$\mathcal A=\bar\rho+Y$, where $Y\subset X$ is a closed linear
subspace, we identify $\mathcal A$ with $Y$ by translation and define
\[
 g(\zeta)=G(\bar\rho+\zeta),\qquad \zeta\in Y.
\]
The reference point $\bar\rho\in\mathcal A$ is arbitrary: replacing it by
$\bar\rho+\zeta$ with $\zeta\in Y$ replaces $g$ by $g(\cdot+\zeta)$, and none of the
quantities or estimates below depends on this choice.
The affine slice $\mathcal A$ allows the regularization and subsequent
minimization to respect linear constraints (such as fixed mass,
see below). The unconstrained case is recovered by taking $\mathcal A=X$.
We observe that $(Y,\norm{\cdot}_Y)$ is a Banach space and $g:Y\to[0,+\infty]$ is
proper and convex.  With this identification, \eqref{eq:restricted-envelope}
reads
\begin{equation}\label{eq:translated-envelope}
 (G+\iota_{\mathcal A})^{\epsilon}(\bar\rho+\zeta)
 =\inf_{\xi\in Y}
 \left\{g(\xi)+\frac{1}{p\epsilon}\norm{\zeta-\xi}_Y^p\right\}
 =g^{\epsilon}(\zeta),
\end{equation}
where $\norm{\cdot}_Y$ denotes the restriction of the chosen regularizing
norm to $Y$.
For the penalty restricted to $Y$, write
\[
 \phi_{\epsilon,Y}=\phi_{\epsilon}|_Y,
 \qquad
 \phi_{\epsilon,Y}^*(\eta)
 =\frac{\epsilon^{q-1}}q\norm{\eta}_{Y^*}^q,
 \quad \eta\in Y^*.
\]
In particular, if the corresponding dual norm on $Y^*$ is
strictly convex, then the envelope $g^{\epsilon}$ on $Y$ is G\^ateaux
differentiable on $Y$.

\subsection{An abstract error estimate}\label{sec:abstract-error}

Let $\mathcal A=\bar\rho+Y$ be a closed affine subspace of $X$ and let $G:X\to[0,+\infty]$ be proper and convex with
$\mathcal A\cap\dom G\ne\varnothing$. Recall that $\bar\rho\in \mathcal A$ is arbitrary. Let $g$ and $g^{\epsilon}$ be as in
\eqref{eq:translated-envelope}.
Define
\[
 Y^\perp=\{w\in X^*:w|_Y=0\}.
\]
For $v\in X^*$, let $[v]=v+Y^\perp\in X^*/Y^\perp$.  The canonical isometric
identification $X^*/Y^\perp\simeq Y^*$ sends $[v]$ to $v|_Y$, and
\begin{equation}\label{eq:quotient-norm}
 \norm{[v]}_{Y^*}
 =\norm{v|_Y}_{Y^*}
 =\inf_{w\in Y^\perp}\norm{v+w}_{X^*},
\end{equation}
where $Y^*$ carries the dual of the regularizing norm $\norm{\cdot}_Y$.
Following the concave-conjugate notation of
\cite{kvaal2014differentiable}, write
$\Phi^{\wedge}(v)=\inf_{\rho\in X}\{\Phi(\rho)+\dua{v}{\rho}\}$ and define (abstract) energies
\begin{align*}
 E_{\mathcal A}(v)
 &=(G+\iota_{\mathcal A})^{\wedge}(v)
  =\inf_{\rho\in\mathcal A}
   \bigl\{G(\rho)+\dua{v}{\rho}\bigr\},
 \\
 E_{\mathcal A}^{\epsilon}(v)
 &=\bigl((G+\iota_{\mathcal A})^{\epsilon}
            +\iota_{\mathcal A}\bigr)^{\wedge}(v)\\
 &=\inf_{\zeta\in Y}
   \bigl\{g^{\epsilon}(\zeta)+
          \dua{v}{\bar\rho+\zeta}\bigr\}.
\end{align*}

In the affine-restricted setting, the usual shift rule for the concave
conjugate of the $p$-MY regularization takes the following form.

\begin{proposition}
\label{prop:exact-affine-shift}
For every $v\in X^*$, as an extended-real identity,
\begin{equation}\label{regeneshift}
 E_{\mathcal A}^{\epsilon}(v)
 =E_{\mathcal A}(v)
  -\frac{\epsilon^{q-1}}{q}\norm{[v]}_{Y^*}^q.
\end{equation}
Moreover, if $w\in Y^\perp$, then
\begin{align*}
 E_{\mathcal A}(v+w)
 &=E_{\mathcal A}(v)+\dua{w}{\bar\rho},\\
 E_{\mathcal A}^{\epsilon}(v+w)
 &=E_{\mathcal A}^{\epsilon}(v)
   +\dua{w}{\bar\rho}.
\end{align*}
\end{proposition}

Let $K\subset\mathcal A$ be compact, and write
$K_Y=K-\bar\rho =\{\sigma-\bar\rho:\sigma\in K\}$.  Let $\iota_K$ denote the indicator of $K$.  Define the
$K$-restricted regularized energy and the associated localization error,
respectively, by
\begin{align}
 E_K^{\epsilon}(v)
 &=\bigl((G+\iota_{\mathcal A})^{\epsilon}
            +\iota_K\bigr)^{\wedge}(v)\notag\\
 &=\min_{\bar\rho+\zeta\in K}
   \bigl\{g^{\epsilon}(\zeta)+
          \dua{v}{\bar\rho+\zeta}\bigr\},
 \notag\\
 \tau_{\mathcal A,K}^{\epsilon}(v)
 &=E_K^{\epsilon}(v)-E_{\mathcal A}^{\epsilon}(v)\ge0.
 \label{eq:localization-error}
\end{align}
The $K$-restricted regularized energy is finite, so
$\tau_{\mathcal A,K}^{\epsilon}(v)\in[0,+\infty]$.
By Proposition~\ref{prop:exact-affine-shift}, this localization
error is finite whenever $E_{\mathcal A}(v)>-\infty$.
Here the comparison concerns the exact regularized functional,
before introducing any network approximation.
Figure~\ref{fig:convex-lower-network} illustrates that the unrestricted network energy can equal
$-\infty$ even when the exact energy is finite.

We now apply the convex approximation of Theorem~\ref{thm:UAPconvex}
to the continuous convex envelope $g^{\epsilon}$.  Combining this
approximation with the exact affine energy shift gives the following
estimate.

\begin{theorem}
\label{thm:ErrorLegendre}
For every $\delta>0$, there exists a finite-rank convex nonnegative
network $\NN:Y\to[0,+\infty)$ such that
\[
 0\le\NN\le g^{\epsilon}\quad\text{on }Y,
 \qquad
 0\le g^{\epsilon}-\NN<\delta\quad\text{on }K_Y.
\]
For any continuous network $\NN$ satisfying these bounds, define its lift to
$\mathcal A$ by
\[
 \widetilde{\NN}(\bar\rho+\zeta)=\NN(\zeta),\qquad \zeta\in Y.
\]
Define its $K$-restricted energy by
\begin{equation*}
 E_{\NN,K}(v)
 =\min_{\bar\rho+\zeta\in K}
 \bigl\{\NN(\zeta)+\dua{v}{\bar\rho+\zeta}\bigr\}.
\end{equation*}
Then every $v\in X^*$ for which $E_{\mathcal A}(v)>-\infty$
satisfies
\begin{equation}\label{eq:absolute-energy-bound}
 \left|E_{\mathcal A}(v)-E_{\NN,K}(v)
       -\frac{\epsilon^{q-1}}{q}\norm{[v]}_{Y^*}^q\right|
 \le\tau_{\mathcal A,K}^{\epsilon}(v)+\delta.
\end{equation}
\end{theorem}

Thus, approximating $g^{\epsilon}$ uniformly from below on $K_Y$ by a
continuous convex network yields a $K$-restricted energy approximation
whose error relative to $E_{\mathcal A}(v)$, after adding the exact
$p$-MY regularization correction to the network energy, is controlled by
the localization error and the network tolerance $\delta$.
The bound also applies to the non-affine convex feature maps
described after Theorem~\ref{thm:UAPconvex}.

\section{Application to density-functional theory}
\label{sec:dft-application}

We now specialize the abstract construction from
Section~\ref{sec:abstract-framework} to electronic density-functional
theory (DFT).  The Banach space becomes the quasidensity space, the affine
restriction fixes the mass, and the quotient of the potential space expresses
the freedom to add a constant to the external potential.  We recall only the
facts used below. The interested reader is referred to broader mathematical accounts, e.g.
\cite{lieb1983density,lewin2019universal,helgakerlieb,Penz2023Structure}.

\subsection{Universal functionals and fixed-mass geometry}
\label{sec:dft-functionals}

Fix an integer particle number $N\ge1$ and, for simplicity, consider spinless
fermions in three dimensions, with complex quantum state spaces.
Following~\cite{lieb1983density}, the natural dual pair of real Banach spaces for densities and potentials is
\begin{equation}\label{eq:dft-dual-pair}
 X=L^1(\RR^3)\cap L^3(\RR^3),
 \qquad
 X^*=L^\infty(\RR^3)+L^{3/2}(\RR^3),
 \qquad
 \dua{v}{\rho}=\int_{\RR^3}v\rho.
\end{equation}
The standard norms are
\begin{equation}\label{eq:dft-natural-norms}
\begin{aligned}
 \norm{\rho}_X
 &=\max\{\norm{\rho}_{L^1},\norm{\rho}_{L^3}\},\\
 \norm{v}_{X^*}
 &=\inf_{\substack{v=v_\infty+v_{3/2}\\
                    v_\infty\in L^\infty(\RR^3),\,
                    v_{3/2}\in L^{3/2}(\RR^3)}}
   \bigl\{\norm{v_\infty}_{L^\infty}
          +\norm{v_{3/2}}_{L^{3/2}}\bigr\}.
\end{aligned}
\end{equation}
In particular, $X$ is separable and its continuous dual is the space in
\eqref{eq:dft-dual-pair}.

Writing $\mathbf x=(x_1,\ldots,x_N)$, let $q_N^0$ be the nonnegative closed
quadratic form on $H_a^1(\RR^{3N})$ given by
\[
 q_N^0[\Psi]
 =\sum_{j=1}^N\int_{\RR^{3N}}|\nabla_{x_j}\Psi(\mathbf x)|^2
    \,\dd\mathbf x
  +\sum_{1\le j<k\le N}\int_{\RR^{3N}}
    \frac{|\Psi(\mathbf x)|^2}{|x_j-x_k|}\,\dd\mathbf x.
\]
For $v\in X^*$, the external-potential form
\[
 V[\Psi]
 =\sum_{j=1}^N\int_{\RR^{3N}}v(x_j)|\Psi|^2
   \,\dd\mathbf x
\]
is infinitesimally form bounded with respect to $q_N^0$, i.e., for every $\eta>0$
there is a constant $C_{\eta,v}$ such that
\[
 |V[\Psi]|
 \le\eta q_N^0[\Psi]+C_{\eta,v}\norm{\Psi}_{L^2}^2,
 \qquad \Psi\in H_a^1(\RR^{3N}).
\]
The KLMN theorem \cite[Theorem~X.17]{reed1975ii} therefore implies that
$q_N^v=q_N^0+V$ is closed and lower semibounded on $H_a^1(\RR^{3N})$ and
determines a unique self-adjoint operator $H_N^v$.  In formal operator
notation,
\begin{equation}\label{eq:dft-hamiltonian}
 H_N^v=H_N^0+\sum_{j=1}^N v(x_j),
 \qquad
 H_N^0=-\sum_{j=1}^N\Delta_{x_j}
       +\sum_{1\le j<k\le N}\frac1{|x_j-x_k|}.
\end{equation}
Throughout the canonical discussion, the particle number $N$ is fixed.
We suppress its subscript on density functionals, energies, and associated
approximation and representability quantities.  Particle spaces,
Hamiltonians, and fixed-mass sets retain their particle-number labels,
as do explicit comparisons between particle-number sectors.
The ground-state energy is
\begin{equation}\label{eq:canonical-ground-state-energy}
 E(v)=\inf_{\substack{\Psi\in H_a^1(\RR^{3N})\\
                         \norm{\Psi}_{L^2}=1}}
        \langle\Psi,H_N^v\Psi\rangle,
\end{equation}
where $\langle\Psi,H_N^v\Psi\rangle=q_N^v[\Psi]$ in the sense of quadratic
forms.
Lieb's analysis \cite[Theorem~3.1]{lieb1983density} shows that
$E:X^*\to\RR$ is concave, locally Lipschitz continuous, and order
preserving: if $v_1\le v_2$ almost everywhere, then
$E(v_1)\le E(v_2)$.  The infimum in
\eqref{eq:canonical-ground-state-energy} need not be attained.
If $\rho_\Psi$ denotes the one-particle density of $\Psi$, then the external
term in \eqref{eq:dft-hamiltonian} is $\dua{v}{\rho_\Psi}$.  Following Lieb's formulation \cite{lieb1983density}, define the canonical mixed-state
constrained-search functional by
\begin{equation*}
 F_{\mathrm L}(\rho)
 =\inf_{\substack{\Gamma\in\mathfrak D_N\\
                    \rho_\Gamma=\rho}}
      \Tr(H_N^0\Gamma),
 \qquad \rho\in X,
\end{equation*}
where $\mathfrak D_N$ is the set of normalized positive $N$-particle density
matrices with finite kinetic energy, and the infimum of the empty set is
$+\infty$.  The standard constrained-search results imply that
$F_{\mathrm L}:X\to[0,+\infty]$ is proper, convex, and norm lower
semicontinuous, with effective domain
\begin{equation}\label{eq:canonical-density-domain}
 \mathcal I_N
 =\left\{\rho\in X:\rho\ge0,\ \sqrt\rho\in H^1(\RR^3),\
                    \int_{\RR^3}\rho=N\right\}.
\end{equation}
Moreover, the Hoffmann--Ostenhof bound \cite{hoffmann1977schrodinger} gives
\begin{equation*}
 F_{\mathrm L}(\rho)\ge \int_{\RR^3}|\nabla\sqrt\rho|^2,
 \qquad \rho\in\mathcal I_N.
\end{equation*}
The Lieb variational principle states that
\begin{equation}\label{eq:lieb-variational-principle}
 E(v)=\inf_{\rho\in X}\bigl\{F_{\mathrm L}(\rho)+\dua{v}{\rho}\bigr\},
 \qquad v\in X^*.
\end{equation}
These facts, including the representability statement in
\eqref{eq:canonical-density-domain}, are
reviewed in \cite{lieb1983density,lewin2019universal}.

The total-mass functional $m\in X^*$, defined by
$m(\rho)=\int_{\RR^3}\rho$, is continuous.  Let us define
\begin{equation}\label{eq:dft-affine-slice}
 \mathcal A_\nu=m^{-1}(\nu)=\bar\rho+Y,
 \qquad
 Y=\ker m,
\end{equation}
for any fixed $\bar\rho\geq 0$ with $\sqrt{\rho}\in H^1$ and $\int_{\RR^3} \rho=\nu$.    Since $m(\rho)=\dua{1}{\rho}$,
\begin{equation}\label{eq:dft-annihilator}
 Y^\perp=\Span\{1\}.
\end{equation}
Thus potentials restricted to $Y$ are potentials modulo additive constants,
and \eqref{eq:lieb-variational-principle} implies the shift identity
\begin{equation*}
 E(v+c)=E(v)+cN,
 \qquad v\in X^*,\quad c\in\RR.
\end{equation*}

We also need the grand-canonical functional.  Following
\cite[Section~2.3]{lewin2019universal}, let
$\mathfrak D_{\mathrm{GC}}$ denote the normalized, block-diagonal states on
fermionic Fock space with finite expected particle number and finite total
kinetic energy.
Here $H_n^0$ is the internal Hamiltonian in \eqref{eq:dft-hamiltonian},
and the density of
$\Gamma=(\Gamma_n)_{n\ge0}$ is $\rho_\Gamma=\sum_{n\ge1}\rho_{\Gamma_n}$.
The grand-canonical universal functional is
\begin{equation}\label{eq:grand-canonical-functional}
 F_{\mathrm{GC}}(\rho)
 =\inf_{\substack{\Gamma=(\Gamma_n)_{n\ge0}
                     \in\mathfrak D_{\mathrm{GC}}\\
                    \rho_\Gamma=\rho}}
   \sum_{n\ge1}\Tr(H_n^0\Gamma_n),
 \qquad \rho\in X.
\end{equation}
The series in \eqref{eq:grand-canonical-functional} is understood in
$[0,+\infty]$.
The functional $F_{\mathrm{GC}}$ is proper, nonnegative, convex, and norm lower semicontinuous. Its
effective domain is
\begin{equation*}
 \mathcal I
 =\{\rho\in X:\rho\ge0,\ \sqrt\rho\in H^1(\RR^3)\}.
\end{equation*}
The Hoffmann--Ostenhof inequality gives
\begin{equation}\label{eq:dft-ho-grand-canonical}
 F_{\mathrm{GC}}(\rho)
 \ge\int_{\RR^3}|\nabla\sqrt\rho|^2,
 \qquad \rho\in\mathcal I.
\end{equation}
For $\nu>0$, the corresponding fixed-mass energy is
\begin{equation*}
 E_\nu^{\mathrm{GC}}(v)
 =\inf_{\rho\in\mathcal A_\nu}
   \bigl\{F_{\mathrm{GC}}(\rho)+\dua{v}{\rho}\bigr\}.
\end{equation*}
We refer to \cite{lewin2019universal} for more details about the grand-canonical functional.

At integer mean mass, every canonical state is an admissible grand-canonical
state, so
\begin{equation*}
 E_N^{\mathrm{GC}}(v)\le E(v),\qquad v\in X^*.
\end{equation*}
Equality need not hold, since fixing the mean particle number still permits
mixtures of different particle-number sectors, see \cite{DiMarino2024}.
We therefore distinguish canonical and fixed-mean-mass grand-canonical
energies even at integer mass.

\subsection{Canonical regularization and neural-network energies}
\label{sec:dft-energy-approximation}

The following result makes the construction in \cite{lammert2007differentiability} rigorous.

\begin{proposition}
\label{prop:dft-local-blowup}
For every $\rho\in\mathcal I_N$, there is a sequence
$(\rho_j)_{j\ge1}\subset\mathcal I_N$ such that
\begin{align*}
 \norm{\rho_j-\rho}_{L^1}
 &\le\frac{N}{2j},
 &
 \norm{\rho_j-\rho}_{L^3}
 &\le\frac{\norm{\rho}_{L^3}}{2j},
\end{align*}
and
\begin{equation*}
 F_{\mathrm L}(\rho_j)
 \ge F_{\mathrm{GC}}(\rho_j)
 \ge\int_{\RR^3}|\nabla\sqrt{\rho_j}|^2
 \longrightarrow+\infty.
\end{equation*}
Consequently, $\rho_j\to\rho$ in $X$, whereas both universal functionals
diverge along the sequence.  In particular, their restrictions to
$\mathcal I_N$ are discontinuous at every point.
\end{proposition}

The unregularized functionals
come with certain challenges that prohibit the use of
Theorems~\ref{uap}, \ref{sigapprox}, and~\ref{thm:UAPconvex}:
their effective domains are not the whole space, and by
Proposition~\ref{prop:dft-local-blowup} even their restrictions to
$\mathcal I_N$ are discontinuous at every point.  We therefore regularize them.
For the regularization, either retain the norm in
\eqref{eq:dft-natural-norms} or replace it by an equivalent norm.  From this
point on, $\norm{\cdot}_X$ denotes the chosen norm and
$\norm{\cdot}_{X^*}$ its dual norm.
Equivalent norms leave the density and potential spaces, their topology, and
the compact subsets of $X$ unchanged, but they change the numerical
regularization penalty.  The restriction of $\norm{\cdot}_X$ to $Y$ is
denoted by $\norm{\cdot}_Y$.  By \eqref{eq:dft-annihilator} and
\eqref{eq:quotient-norm},
\begin{equation}\label{eq:dft-quotient-norm}
 \norm{[v]}_{Y^*}
 =\inf_{c\in\RR}\norm{v+c}_{X^*}.
\end{equation}

The fixed-mass $p$-MY regularization is the restriction to $\mathcal A_N$ of
the ordinary envelope on $X$:
\begin{equation}\label{eq:dft-restricted-envelope}
 F_{\mathrm L}^{\epsilon}(\rho)
 =\inf_{\sigma\in X}
   \left\{F_{\mathrm L}(\sigma)
       +\frac1{p\epsilon}\norm{\rho-\sigma}_X^p\right\},
 \qquad \rho\in\mathcal A_N.
\end{equation}
Indeed, $\dom F_{\mathrm L}\subset\mathcal A_N$, so $F_{\mathrm L}(\sigma)=+\infty$
unless $\sigma\in\mathcal A_N$, and the infimum in
\eqref{eq:dft-restricted-envelope} may be taken over $\sigma\in\mathcal A_N$,
as in \eqref{eq:restricted-envelope}.  To express the
functional on the Banach space $Y$ and apply the preceding results, set
\begin{equation*}
 g(\zeta)=F_{\mathrm L}(\bar\rho+\zeta),
 \qquad \zeta\in Y.
\end{equation*}
Here $\bar\rho$ is the reference density from \eqref{eq:dft-affine-slice}.
By \eqref{eq:translated-envelope},
$F_{\mathrm L}^{\epsilon}(\bar\rho+\zeta)=g^{\epsilon}(\zeta)$.
The corresponding regularized energy is
\begin{equation*}
 E^{\epsilon}(v)
 =\inf_{\rho\in\mathcal A_N}
   \bigl\{F_{\mathrm L}^{\epsilon}(\rho)+\dua{v}{\rho}\bigr\}.
\end{equation*}
For $\epsilon>0$, the conjugate of the restricted penalty gives the
nonnegative magnitude of the downward regularization shift
\begin{equation}\label{eq:dft-regularization-correction}
 \phi_{\epsilon,Y}^*([v])
 =\frac{\epsilon^{q-1}}q\norm{[v]}_{Y^*}^q
 =\frac{\epsilon^{q-1}}q
   \left(\inf_{c\in\RR}\norm{v+c\,1}_{X^*}\right)^q,
\end{equation}
so that Proposition~\ref{prop:exact-affine-shift} now reads
\begin{equation}\label{eq:dft-regularized-shift}
 E^{\epsilon}(v)=E(v)-\phi_{\epsilon,Y}^*([v]).
\end{equation}
Adding this regularization correction to $E^{\epsilon}(v)$ therefore
recovers $E(v)$ exactly.
After translation by $\bar\rho$, the envelope in
\eqref{eq:dft-restricted-envelope} is finite, convex, and locally Lipschitz on
$Y$.

Let $K\subset\mathcal I_N$ be compact.  The localization error is
\begin{equation*}
\tau_K^{\epsilon}(v)
 =\min_{\rho\in K}
   \bigl\{F_{\mathrm L}^{\epsilon}(\rho)+\dua{v}{\rho}\bigr\}
 -E^{\epsilon}(v)\ge0.
\end{equation*}

All hypotheses of Theorem~\ref{thm:ErrorLegendre} are now satisfied. Its
regularization correction depends only on the potential modulo constants,
and the remaining errors come from restriction to the density set $K$ and
neural-network approximation.

\begin{corollary}
\label{cor:dft-energy-approximation}
Write
$K_Y=K-\bar\rho\subset Y$ for the
translate of $K$ into $Y$.
For every $\delta>0$, there exists a finite-rank convex nonnegative
network $\NN:Y\to[0,+\infty)$ such that
\begin{equation}\label{eq:dft-network-approximation}
 0\le\NN\le g^{\epsilon}\quad\text{on }Y,
 \qquad
 0\le g^{\epsilon}-\NN<\delta\quad\text{on }K_Y.
\end{equation}
For any continuous network satisfying \eqref{eq:dft-network-approximation},
define its $K$-restricted ground-state energy by
\begin{equation}\label{eq:dft-network-energy}
 E_{\NN,K}(v)
 =\min_{\bar\rho+\zeta\in K}
   \bigl\{\NN(\zeta)+\dua{v}{\bar\rho+\zeta}\bigr\}.
\end{equation}
Then every $v\in X^*$ satisfies
\begin{equation}\label{eq:dft-absolute-energy-estimate}
 \left|E(v)-E_{\NN,K}(v)-\phi_{\epsilon,Y}^*([v])\right|
 \le\tau_K^{\epsilon}(v)+\delta.
\end{equation}
\end{corollary}

Thus adding the potential-dependent regularization correction to the network
energy leaves an error controlled by the localization error determined by
$K$ and the network tolerance.

\begin{proposition}
\label{prop:dft-network-energy}
Let $\NN$ and $K$ be as in Corollary~\ref{cor:dft-energy-approximation}.
Then $E_{\NN,K}:X^*\to\RR$ is finite and concave, and Lipschitz
\begin{equation}\label{eq:dft-network-energy-lipschitz}
 |E_{\NN,K}(v)-E_{\NN,K}(w)|
 \le\Bigl(\max_{\rho\in K}\norm{\rho}_X\Bigr)\norm{v-w}_{X^*},
 \qquad v,w\in X^*.
\end{equation}
Moreover,
\begin{equation}\label{eq:dft-network-energy-shift}
 E_{\NN,K}(v+c)=E_{\NN,K}(v)+cN,
\end{equation}
and $E_{\NN,K}(v)\le E_{\NN,K}(w)$ whenever $v\le w$ almost everywhere.
\end{proposition}

\subsection{Grand-canonical regularization and neural-network energies}
\label{sec:dft-gc-energy-approximation}

Fix $\nu>0$, put $\mathcal I_\nu=\mathcal I\cap\mathcal A_\nu$, and set
$\bar\rho_\nu=(\nu/N)\bar\rho\in\mathcal I_\nu$.  Then
$\mathcal A_\nu=\bar\rho_\nu+Y$.  Unlike $F_{\mathrm L}$, the effective
domain of $F_{\mathrm{GC}}$ is not contained in a single mass slice.
Define
\begin{equation*}
 g_\nu^{\mathrm{GC}}(\zeta)
 =F_{\mathrm{GC}}(\bar\rho_\nu+\zeta),
 \qquad \zeta\in Y,
\end{equation*}
and
\begin{equation}\label{eq:dft-gc-restricted-envelope}
 F_{\mathrm{GC},\nu}^{\epsilon}(\rho)
 :=\inf_{\sigma\in\mathcal A_\nu}
   \left\{F_{\mathrm{GC}}(\sigma)
       +\frac1{p\epsilon}\norm{\rho-\sigma}_X^p\right\},
 \qquad \rho\in\mathcal A_\nu.
\end{equation}
Under translation by $\bar\rho_\nu$, this becomes
\begin{equation*}
 F_{\mathrm{GC},\nu}^{\epsilon}(\bar\rho_\nu+\zeta)
 =\bigl(g_\nu^{\mathrm{GC}}\bigr)^{\epsilon}(\zeta)
 =\inf_{\xi\in Y}
   \left\{F_{\mathrm{GC}}(\bar\rho_\nu+\xi)
       +\frac1{p\epsilon}\norm{\zeta-\xi}_Y^p\right\}.
\end{equation*}
In general, \eqref{eq:dft-gc-restricted-envelope} differs from
$F_{\mathrm{GC}}^{\epsilon}|_{\mathcal A_\nu}$, since the latter allows
competitors of masses other than $\nu$.

The regularized fixed-mass energy is
\begin{equation*}
 E_\nu^{\mathrm{GC},\epsilon}(v)
 =\inf_{\rho\in\mathcal A_\nu}
   \bigl\{F_{\mathrm{GC},\nu}^{\epsilon}(\rho)
          +\dua{v}{\rho}\bigr\}.
\end{equation*}
Since $E_\nu^{\mathrm{GC}}(v)$ is finite for every $v\in X^*$,
Proposition~\ref{prop:exact-affine-shift} gives
\begin{equation}\label{eq:dft-gc-regularized-shift}
 E_\nu^{\mathrm{GC},\epsilon}(v)
 =E_\nu^{\mathrm{GC}}(v)-\phi_{\epsilon,Y}^*([v]).
\end{equation}
Here $\phi_{\epsilon,Y}^*([v])$ is the same regularization correction as in
\eqref{eq:dft-regularization-correction}.  The exact energy
$E_\nu^{\mathrm{GC}}$ and the regularized energy
$E_\nu^{\mathrm{GC},\epsilon}$ both shift by $c\nu$ under
$v\mapsto v+c$.

Let $K\subset\mathcal I_\nu$ be compact.  Define the
grand-canonical localization error by
\begin{equation*}
 \tau_{\nu,K}^{\mathrm{GC},\epsilon}(v)
 =\min_{\rho\in K}
   \bigl\{F_{\mathrm{GC},\nu}^{\epsilon}(\rho)+\dua{v}{\rho}\bigr\}
  -E_\nu^{\mathrm{GC},\epsilon}(v)\ge0.
\end{equation*}
Since $E_\nu^{\mathrm{GC}}$ is finite, Theorem~\ref{thm:ErrorLegendre}
with $G=F_{\mathrm{GC}}$ and $\mathcal A=\mathcal A_\nu$ applies to every
$v\in X^*$ and gives the grand-canonical analogue of
Corollary~\ref{cor:dft-energy-approximation}.

\begin{corollary}
\label{cor:dft-gc-energy-approximation}
Write
$K_{\nu,Y}:=K-\bar\rho_\nu
\subset Y$ for the translate of $K$
into $Y$.
For every $\delta>0$, there exists a finite-rank convex nonnegative
network $\NN_\nu:Y\to[0,+\infty)$ such that
\begin{equation}\label{eq:dft-gc-network-approximation}
 0\le\NN_\nu\le\bigl(g_\nu^{\mathrm{GC}}\bigr)^{\epsilon}
 \quad\text{on }Y,
 \qquad
 0\le\bigl(g_\nu^{\mathrm{GC}}\bigr)^{\epsilon}-\NN_\nu<\delta
 \quad\text{on }K_{\nu,Y}.
\end{equation}
For any continuous network satisfying \eqref{eq:dft-gc-network-approximation},
define its $K$-restricted energy by
\begin{equation*}
 E_{\NN_\nu,K}^{\mathrm{GC}}(v)
 =\min_{\bar\rho_\nu+\zeta\in K}
   \bigl\{\NN_\nu(\zeta)+\dua{v}{\bar\rho_\nu+\zeta}\bigr\}.
\end{equation*}
Then every $v\in X^*$ satisfies
\begin{equation}\label{eq:dft-gc-absolute-energy-estimate}
 \left|E_\nu^{\mathrm{GC}}(v)-E_{\NN_\nu,K}^{\mathrm{GC}}(v)
       -\phi_{\epsilon,Y}^*([v])\right|
 \le\tau_{\nu,K}^{\mathrm{GC},\epsilon}(v)+\delta.
\end{equation}
\end{corollary}

\subsection{Localization and uniform energy approximation}
\label{sec:dft-uniform-energy-approximation}

We now return to the canonical setting. The grand-canonical case is
treated in the remark following
Proposition~\ref{prop:dft-compact-potential-classes}.
A fixed compact density set can prevent energy convergence even when the
regularization shift and network approximation error vanish.  Indeed, for
$v\equiv 0$ we have $E(0)=0$, whereas every
fixed compact $K\subset\mathcal I_N$ satisfies (see
Section~\ref{proofsec} for the proof)
\begin{equation}\label{eq:dft-zero-potential-localization}
 0<\tau_K^{\epsilon}(0)
   =\min_{\rho\in K}F_{\mathrm L}^{\epsilon}(\rho)
   \nearrow\min_{\rho\in K}F_{\mathrm L}(\rho)>0
   \qquad(\epsilon\downarrow0).
\end{equation}
The zero energy is approached by increasingly diffuse densities, which a
fixed compact set cannot capture.  The same obstruction occurs in the
grand-canonical setting at fixed mean mass.  The following proposition
therefore allows the localization set to depend on the desired accuracy.

\begin{proposition}
\label{prop:dft-compact-potential-classes}
Let $\mathcal V\subset X^*$ be nonempty and compact in the dual norm.
For every $\eta>0$, there is a compact convex set
$K_\eta\subset\mathcal I_N$, independent of $\epsilon$ and $\delta$, such
that
\begin{equation}\label{eq:dft-compact-class-localization}
 \left|\min_{\rho\in K_\eta}
       \{F_{\mathrm L}(\rho)+\dua{v}{\rho}\}-E(v)\right|
   \le\eta,
 \qquad v\in\mathcal V.
\end{equation}
For every $\epsilon,\delta>0$, there is a network as in
Corollary~\ref{cor:dft-energy-approximation} on $K_\eta$ whose energy satisfies
\begin{equation}\label{eq:dft-compact-class-energy-bound}
 \sup_{v\in\mathcal V}|E(v)-E_{\NN,K_\eta}(v)|
 \le\max\left\{\eta,
       \frac{\epsilon^{q-1}}q
       \max_{v\in\mathcal V}\norm{[v]}_{Y^*}^q+\delta
       \right\}.
\end{equation}
Taking $\epsilon,\delta,\eta\to0$, with the localization sets and networks
chosen accordingly, yields uniform energy convergence on $\mathcal V$.
\end{proposition}

\begin{remark}
Neither ground-state attainment nor continuity of ground-state densities
is required.  The same conclusions hold at fixed mean mass $\nu>0$ with
$F_{\mathrm L}$ and $E$ replaced by $F_{\mathrm{GC}}$ and
$E_\nu^{\mathrm{GC}}$, with localization sets $K_\eta\subset\mathcal I_\nu$
and the network energies of Corollary~\ref{cor:dft-gc-energy-approximation}.
\end{remark}

Proposition~\ref{prop:dft-compact-potential-classes} applies to families of
three-dimensional molecular Coulomb potentials.  Fix a finite number $J$ of nuclei with charges
$Z_1,\ldots,Z_J>0$ and let their positions
$R=(R_1,\ldots,R_J)$ range over a nonempty compact set
$\mathcal R\subset(\RR^3)^J$.  The corresponding electronic external
potentials are
\begin{equation*}
 v_R(x)=-\sum_{a=1}^J\frac{Z_a}{|x-R_a|},
 \qquad
 \mathcal V_{\mathcal R}=\{v_R:R\in\mathcal R\}.
\end{equation*}
The map $R\mapsto v_R$ is norm continuous into
$L^\infty(\RR^3)+L^{3/2}(\RR^3)$, so $\mathcal V_{\mathcal R}$ is
norm compact.  Indeed, a smooth cutoff decomposes the Coulomb kernel into an
$L^{3/2}$ part and a bounded uniformly continuous remainder, whose translations
are norm continuous in $L^{3/2}$ and $L^\infty$, respectively.
This also gives continuity for any equivalent dual norm.

The bound in \eqref{eq:dft-compact-class-energy-bound} therefore holds
uniformly for molecular electronic energies over $\mathcal R$.
If $\mathcal R$ contains no nuclear collisions, the
nuclear-repulsion term
\[
 U(R)=\sum_{1\le a<b\le J}\frac{Z_aZ_b}{|R_a-R_b|}
\]
is finite and continuous on $\mathcal R$.  Adding $U(R)$ exactly to the
exact and learned electronic energies leaves their difference unchanged.
The same uniform error bound thus holds for the clamped-nuclei
Born--Oppenheimer energy surface on this compact configuration set.

\section{Numerical experiments}\label{sec:numerical-experiments}

We compare ReLU approximations of a two-fermion density functional and
its regularized version against exact reference energies.
To that end, we consider a four-site instance of
the periodic spinless $t$--$V$ model, with the interaction
strength denoted here by $U$ \cite{gebhard2022spinless}.
The corresponding noninteracting square-graph model is
discussed in \cite[Sec.~III.C]{penz2021graphs}.
Section~\ref{sec:num-results} also compares the affine constructions
with networks trained from functional values by minimizing the
mean-square error with Adam.

\subsection{Model and reference calculations}

We consider two spinless fermions on a ring with four sites, where site indices are understood modulo four. Particles can move
between adjacent sites with hopping amplitude $t$. We express energies
in units of $t$ and set $t=1$ throughout the numerical experiments. We focus on the noninteracting case $U=0$ as well as a nearest-neighbor
repulsion of strength $U=2$.

Let $e_0,\ldots,e_3$ be the standard orthonormal basis of
$\mathbb C^4$. The two-fermion Hilbert space is the antisymmetric
tensor product
\[
 \mathcal H_2=\bigwedge\nolimits^2\mathbb C^4.
\]
Its normalized basis vectors are
\[
 e_{ab}=e_a\wedge e_b
 =\frac{e_a\otimes e_b-e_b\otimes e_a}{\sqrt2},
 \qquad 0\le a<b\le3.
\]
The vector $e_{ab}$ describes a configuration with sites $a$
and $b$ occupied. Antisymmetry excludes double occupation and
identifies exchanging the two particles with a change of sign.
Consequently, $\dim\mathcal H_2=\binom42=6$. We use the ordered
basis
\[
\mathcal B = (e_{01},e_{02},e_{03},e_{12},e_{13},e_{23}).
\]

We take the one-particle kinetic energy to be the discrete Laplacian
$L e_j = -e_{j-1} + 2e_j - e_{j+1}$, and the corresponding two-particle version
$K (e_a \wedge e_b) = (Le_a)\wedge e_b+e_a\wedge(Le_b)$. An external potential
$v=(v_0,v_1,v_2,v_3)\in\mathbb R^4$ contributes the energy
$v_a+v_b$ to the configuration $e_{ab}$. The interaction
contributes $U$ when the occupied sites are adjacent and
zero otherwise, which makes the interaction diagonal in our choice of basis.
In the ordered basis $\mathcal B$, we decompose the Hamiltonian
into kinetic, potential, and interaction contributions,
\[
\begin{aligned}
H_2^{v,U}
={}&
\begin{pmatrix}
4&-1&0&0&1&0\\
-1&4&-1&-1&0&1\\
0&-1&4&0&-1&0\\
0&-1&0&4&-1&0\\
1&0&-1&-1&4&-1\\
0&1&0&0&-1&4
\end{pmatrix}
\\[1ex]
&+
\operatorname{diag}\bigl(
v_0+v_1,\,
v_0+v_2,\,
v_0+v_3,\,
v_1+v_2,\,
v_1+v_3,\,
v_2+v_3
\bigr)
\\[1ex]
&+
U\,\operatorname{diag}(1,0,1,1,0,1).
\end{aligned}
\]

The occupation operator $n_j$ is diagonal in the basis $\mathcal B$,
$n_j e_{ab}=(\delta_{ja}+\delta_{jb})e_{ab}$,
with eigenvalues zero and one, indicating whether
site $j$ is empty or occupied.
An ensemble state is represented by a positive semidefinite
Hermitian $6\times6$ matrix $\Gamma$ with $\Tr\Gamma=1$.
Its density is the vector of mean site occupations,
\begin{equation}\label{eq:Gamma-nj}
 \rho_j=\Tr(n_j\Gamma),\qquad j=0,1,2,3.
\end{equation}
The attainable densities, i.e., those for which there is a state $\Gamma$ fulfilling Eq.~\eqref{eq:Gamma-nj},
are given by
\[
 \mathcal I_2
 =\left\{\rho\in[0,1]^4\ \middle|\ \sum_{j=0}^3\rho_j=2\right\}.
\]

For each fixed $U$, suppressing its occurrence in the notation,
we define the reference functional and ground-state energy by
\[
 F(\rho)
 =\min_{\substack{\Gamma\succeq0,\ \Tr\Gamma=1\\
                   \Tr(n_j\Gamma)=\rho_j,\ j=0,\ldots,3}}
       \Tr(H_2^{0,U}\Gamma),
 \qquad \rho\in\mathcal I_2,
\]
and
\[
 E(v)=\lambda_{\min}(H_2^{v,U})
     =\min_{\rho\in\mathcal I_2}
        \left\{F(\rho)+\sum_{j=0}^3v_j\rho_j\right\}.
\]
We set $F=+\infty$ outside $\mathcal I_2$.
Evaluating $F$ on $\mathcal I_2$ is therefore a semidefinite optimization problem
with linear constraints, whereas evaluating $E$ requires only
the smallest eigenvalue of a $6\times6$ matrix.

To connect this model with the preceding framework, we set
\[
 \bar\rho=\tfrac12(1,1,1,1),\qquad
 Y=\left\{ \rho \in\mathbb R^4\ \middle|\ \sum_{j=0}^3 \rho_j=0\right\},
 \qquad
 \mathcal A_2=\bar\rho+Y.
\]
The Euclidean pairing identifies $Y^*$ with $Y$, so we represent potentials
by zero-mean vectors $v\in Y$. Adding a constant $c$ to every component of $v$
adds $2cI$ to the Hamiltonian and hence shifts the ground-state
energy by $2c$.

\begin{remark}
The computations for $U=0$ and $U=2$ use the same
$6\times6$ semidefinite formulation
for two fermions on four sites.
For larger systems, the noninteracting case permits an exact
reduction to a one-particle density-matrix problem. A nonzero interaction
generally prevents this reduction, making the growth of the full
many-body dimension $\binom{m}{N}$ relevant, where $m$ and $N$ denote
the numbers of sites and particles.
\end{remark}

\subsection{Finite-rank approximations}
\label{sec:num-networks}
For $p=2$, we set
$F^\epsilon(x)=\min_{\sigma\in\mathcal I_2}
\{F(\sigma)+\norm{x-\sigma}_2^2/(2\epsilon)\}$ on $\mathcal A_2$.
We use $\epsilon=0.05$ throughout the numerical section unless stated
otherwise. For any $a\in Y^*$,
$$
h_a^\epsilon(\rho)=\dua a{\rho-\bar\rho}+E(-a)
                    -\tfrac\epsilon2\norm a_2^2
$$
is an affine lower bound for $F^\epsilon$. We define the ReLU approximation
of $F^\epsilon$ using $M$ affine functions, in addition to the constant $E(0)$, by
\begin{equation*}
 \NN_M^\epsilon=\max\{E(0),h_{a_1}^\epsilon,\ldots,h_{a_M}^\epsilon\}.
\end{equation*}
Here $M$ counts the affine functions in the pointwise maximum,
excluding the constant $E(0)$.

In this setting we choose particular compact sets.
Let $ R>0$ and define
$K_{\epsilon R}=\{\rho\in\mathcal A_2:
\operatorname{dist}(\rho,\mathcal I_2)\le\epsilon R\}$, as well as
\begin{equation}\label{eq:num-corrected-energy}
 \widehat E_M^\epsilon(v)=
 \min_{\rho\in K_{\epsilon R}}\{\NN_M^\epsilon(\rho)+\dua v\rho\}
       +\tfrac\epsilon2\norm v_2^2.
\end{equation}
For $\|v\|\leq R$, if $\sigma_v$ is a ground-state density, $\sigma_v-\epsilon v\in K_{\epsilon R}$
is a regularized minimizer, as follows by completing the square.
Localization error is therefore zero, and the
added quadratic term removes the exact regularization shift.
The ReLU approximation of $F$ uses $h_a^0$ and minimizes on
$\mathcal I_2$, without an energy correction.
We compute the network energies with Clarabel~\cite{goulart2024clarabel}.
The minimization is a linear program for the ReLU approximation of $F$
and a second-order cone program for the ReLU approximation of $F^\epsilon$.

For each interaction strength $U=0,2$, we construct these approximations
by adding affine functions one step at a time,
starting from the constant $E(0)$ and using a fixed pool of $256$ sampled
densities. At each step, we select, among the sampled densities not
previously selected, the density where the reference functional exceeds
the current network by the largest amount. This is the greedy selection rule used below.
The sampled densities lie in $\mathcal I_2$ when approximating $F$ and in
$K_{0.1}$ when approximating $F^\epsilon$. Here $K_{0.1}$ consists of points in
$\mathcal A_2$ at distance at most $0.1=\epsilon R$ from $\mathcal I_2$,
with $R=2$. When fitting $F$ directly, a selected density near the boundary
is moved slightly into the interior before fitting its function, using a
margin of $10^{-4}$. This improves numerical stability, since the supporting
slopes of $F$ can become unbounded near the boundary.
Selection continues to use the original sampled densities and their
reference values.

For the ReLU approximation of $F^\epsilon$, let $x_i$ be the selected
density. Substituting the density-matrix definition of $F$ into the
minimization defining $F^\epsilon(x_i)$ gives
\[
 F^\epsilon(x_i)
 =\min_{\substack{\Gamma\succeq0\\ \Tr\Gamma=1}}
   \left\{\Tr(H_2^{0,U}\Gamma)
          +\frac{\norm{x_i-\rho_\Gamma}_2^2}{2\epsilon}\right\},
 \qquad (\rho_\Gamma)_j=\Tr(n_j\Gamma).
\]
We solve this convex semidefinite problem with a quadratic objective
using Clarabel~\cite{goulart2024clarabel}. The variable $\Gamma$ is a
$6\times6$ density matrix.
The density $\sigma_i^\epsilon=\rho_{\Gamma_i^\epsilon}$ of a minimizing
matrix $\Gamma_i^\epsilon$ is the proximal point and determines the slope
\[
 a_i=\frac{x_i-\sigma_i^\epsilon}{\epsilon}
     =\nabla_Y F^\epsilon(x_i),
\]
where $\nabla_Y$ denotes the gradient along the fixed-mass space.
The same optimization therefore provides the reference value
$F^\epsilon(x_i)$, the proximal density, and the slope $a_i$.
We then add $h_{a_i}^\epsilon$ to the maximum defining the network.
Thus the selected density determines the slope, and the formula for
$h_{a_i}^\epsilon$ fixes the intercept through $E(-a_i)$ and the quadratic
correction.

For the ReLU approximation of $F$, let $\sigma_i$ denote the selected
density after any inward adjustment. We solve the semidefinite program
defining $F(\sigma_i)$ with Clarabel. Along with the minimizing matrix,
the solver returns Lagrange multipliers $z_{i,j}$ for the four density
constraints $\Tr(n_j\Gamma)=\sigma_{i,j}$. For $\Gamma\succeq0$ with
$\Tr\Gamma=1$, the corresponding Lagrangian is
\[
 \begin{aligned}
 \mathcal L(\Gamma,z_i)
 &=\Tr(H_2^{0,U}\Gamma)
   +\sum_{j=0}^3z_{i,j}\bigl(\Tr(n_j\Gamma)-\sigma_{i,j}\bigr)\\
 &=\Tr(H_2^{z_i,U}\Gamma)-\dua{z_i}{\sigma_i}.
 \end{aligned}
\]
Thus the multipliers enter the Hamiltonian as an external potential.
When needed, we refine
the computed multipliers using damped Newton steps to increase the dual
value $E(z_i)-\dua{z_i}{\sigma_i}$.
We then normalize the obtained multipliers to a
zero-mean potential and set $a_i=-v_i$.
We compute $E(v_i)$ from the
smallest eigenvalue of $H_2^{v_i,U}$.

For each function constructed in this way, we also add the three functions
obtained by shifting all site labels by one, two, or three positions
around the ring. These relabellings use the rotational symmetry of the
model and give four functions per selection. After $128$ selections,
the network therefore contains $M=512$ functions in addition to $E(0)$.
We repeat the construction with three fixed random seeds.

We assess energy accuracy and functional accuracy on separate test sets
that are independent of the construction data. Energy tests use $32$
potentials in total, with Euclidean norms chosen from $0.5,1,1.5,2$.
Their four components sum to zero, removing the arbitrary constant
offset in the potential. We compare each approximate energy with the
smallest eigenvalue of $H_2^{v,U}$. For the ReLU approximation of
$F^\epsilon$, the energy includes the correction in
\eqref{eq:num-corrected-energy}.
Functional tests use $128$ independent densities for each target.
The ReLU approximation of $F^\epsilon$ is tested on $K_{0.1}$,
and the ReLU approximation of $F$ on $\mathcal I_2$.

We check reference values used for fitting and testing using the gap between
independently recomputed primal and dual values. The gap targets are
$10^{-9}$ for $F$ and $10^{-8}$ for $F^\epsilon$.
These tolerances concern the optimization objective, rather than
individual density components. For $F^\epsilon$, a calculation that
fails the gap check is retried with adjusted solver settings or an
equivalent formulation.
The gaps in the network-energy minimizations are recorded separately
from the approximation errors reported below. For the affine networks
with $M=512$, these gaps are below $3\times10^{-8}$.

We report absolute energy errors over the $32$ test potentials and
absolute functional errors on the corresponding density test sets.
Each reported quantity is the median over the three random seeds.
The largest sampled functional error need not bound the error
elsewhere in the density domain.

For each fixed construction run, the finite-rank approximations are
nondecreasing in $M$, since new affine functions are added while all
earlier ones are retained. With $\epsilon$ and $R$ fixed, for
$M_1\le M_2$ and $\norm v_2\le R$ we have
\[
 \NN_{M_1}^\epsilon\le\NN_{M_2}^\epsilon\le F^\epsilon,
 \qquad
 \widehat E_{M_1}^\epsilon(v)\le\widehat E_{M_2}^\epsilon(v)\le E(v).
\]
The same ordering holds for the direct approximations of $F$, with
energies minimized over $\mathcal I_2$. Thus functional and energy errors
on fixed test sets are nonincreasing in $M$, up to numerical evaluation
and minimization tolerances, although plateaus remain possible.

\subsection{Numerical results and comparison with trained networks}
\label{sec:num-results}
\label{sec:num-learned}

\begin{figure}[htbp]
\centering
\includegraphics[width=\textwidth]{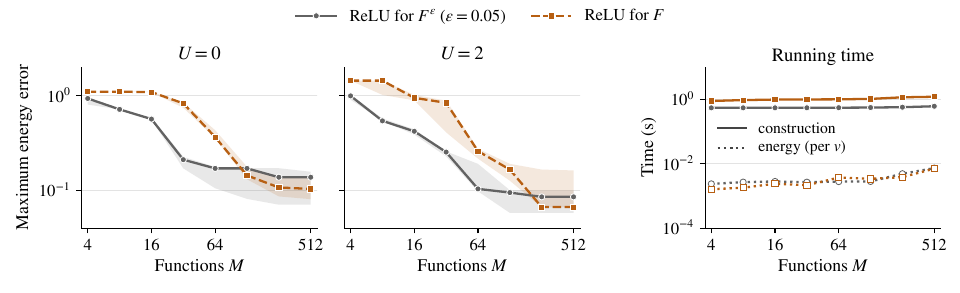}
\caption{The left and center panels show the maximum energy error on
the $32$ test potentials as the number $M$ of affine functions increases,
with $256$ sampled densities, $R=2$, and $\epsilon=0.05$ for $F^\epsilon$.
Energies for $F^\epsilon$ include the correction in
\eqref{eq:num-corrected-energy}.
Lines show medians over three random seeds, and shaded regions show
the range of their results. The right panel shows network construction
time and the time needed to compute one network energy, as medians
over both values of $U$ and the three random seeds. Construction times
include computing reference values for the sampled densities. }
\label{fig:num-energy-error}
\end{figure}

We first examine the affine constructions of
Section~\ref{sec:num-networks}, then compare them with networks trained
from functional values. Figure~\ref{fig:num-energy-error} shows how
the energy errors and running times change as more affine functions
are added, using the fixed pool of $256$ sampled densities.
We observe that the error systematically decreases for both $U=0$ and $U=2$
as well as for the targets $F$ and $F^\epsilon$. Furthermore,
one can see that the interacting case behaves similarly to the noninteracting one.

Computing the reference functional values for the $256$ sampled
densities by semidefinite optimization accounts for most of the
construction time in Figure~\ref{fig:num-energy-error}.
Running times varied substantially between runs, limiting comparisons
between the two interaction strengths. For this small model,
exact diagonalization of $H_2^{v,U}$ remains much faster than computing
a network energy.

\begin{figure}[htbp]
\centering
\includegraphics[width=0.95\textwidth]{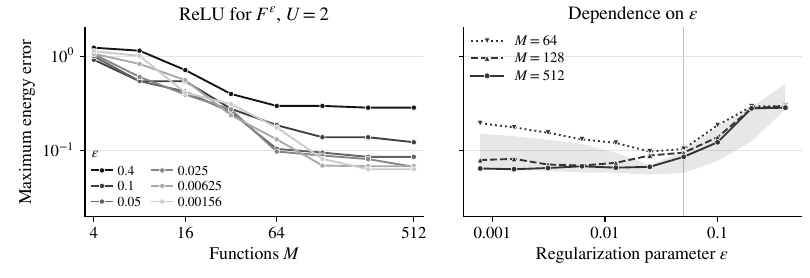}
\caption{ReLU approximations of $F^\epsilon$ for $U=2$ and different values of
$\epsilon$, using the construction in Section~\ref{sec:num-networks}.
The left panel shows the maximum energy error on the $32$ test
potentials as $M$ increases. The right panel shows the same error
as $\epsilon=0.4\cdot2^{-k}$, $k=0,\ldots,9$, varies, with a
separate curve for each fixed $M$. The shaded region for $M=512$
shows the range over three random seeds. The vertical line marks
$\epsilon=0.05$, the value used in Figure~\ref{fig:num-energy-error}.}
\label{fig:num-eps}
\end{figure}

Figure~\ref{fig:num-eps} examines the choice of $\epsilon$ for $U=2$
and $R=2$, using $\epsilon=0.4\cdot2^{-k}$, $k=0,\ldots,9$.
For each value, $K_{\epsilon R}$ is the domain for both density sampling
and energy minimization. At $\epsilon=0.2$ and $0.4$, the median
maximum energy error remains near $0.28$ as $M$ increases from $64$
to $512$, despite including the regularization correction.
This plateau may partly reflect the larger approximation domain at
larger $\epsilon$, over which the same number of $256$ sampled
densities is distributed. Increasing $M$ adds affine functions from
this fixed density pool without adding new densities. However,
larger $\epsilon$ also makes $F^\epsilon$ smoother, so these tests
do not establish the cause of the plateau.

In the preceding tests, the network values and approximate energies were no
larger than their respective reference values. This agrees with the
construction from affine lower bounds and,
for $F^\epsilon$, the vanishing localization error.

Next, we compare the affine constructions with networks fitted to functional
values by minimizing the mean-square error (MSE), using $F$ and
$F^\epsilon$ with $\epsilon=0.05$ as targets for $U=0,2$.
This tests approximation from values without supporting slopes. The multilayer perceptron (MLP) uses $\tanh$ activations and an affine
output. The input-convex neural network (ICNN)~\cite{amos2017input}
uses $s(z)=\log(1+e^z)$. Both have two hidden layers of width
$w=16,32,64$. In the ICNN, weights between hidden layers and from
the last hidden layer to the output are nonnegative, with unrestricted
affine input terms in the second hidden layer and the output.
This ensures convexity in the density, but training remains nonconvex
in the parameters. Neither model is constrained to lie below its target.
Their parameter counts are $w^2+7w+1$ for the MLP and $w^2+11w+5$
for the ICNN.

Both models take $\rho-\bar\rho$ as input and average their outputs
over the four rotations of the site labels around the ring. Sharing
parameters across these evaluations preserves rotational symmetry
without increasing the parameter count. For either target $T$, we minimize
\[
 \frac{1}{5120}\sum_{j=1}^{5120}
 \bigl(\widehat T_\theta(\rho_j)-T(\rho_j)\bigr)^2,
\]
where $\widehat T_\theta$ denotes the averaged network output.

Training uses $5120$ sampled densities for each target,
interaction strength, and random seed, lying in $\mathcal I_2$ for $F$
and $K_{0.1}$ for $F^\epsilon$. Model selection uses $128$ independent
validation densities. For both MLP and ICNN, we run
Adam~\cite{kingma2015adam} for at most $6000$ full-batch
updates at each of the learning rates $0.003$ and $0.01$. We evaluate
the validation MSE every $25$ updates and stop a run early after $750$
updates without improvement. For each learning rate, we retain the
parameters with the lowest observed validation MSE. We then select
between the two saved networks using the smaller validation MSE.
This validation-based model selection is performed separately for
each architecture, width, target, interaction strength, and seed. All choices
are fixed before evaluating the test data.

For both targets, we report functional errors on the same $128$
independent densities in $\mathcal I_2$, always comparing against the
network's own target. This common test set makes the functional errors
easier to compare, although the training domains still differ.
Energy tests use the same $32$ test potentials as in
Figure~\ref{fig:num-energy-error}. We minimize
$\widehat F_\theta(\rho)+\dua v\rho$ over $\mathcal I_2$, or
$\widehat F^\epsilon_\theta(\rho)+\dua v\rho$ over $K_{0.1}$ and add
$\epsilon\norm v_2^2/2$. The latter set contains an exact minimizer
for every test potential. Hence the regularized energy comparison
includes neither an uncorrected energy shift nor a localization error.

We perform two numerical searches for each network energy, each with
eight initial points chosen from a separately sampled set of densities.
Both searches use sequential quadratic programming with analytic
gradients, and we retain the smaller energy. For the ICNN, convexity
also gives a lower bound from the tangent plane at the computed density.
We use the gap between this bound and the computed energy to check the
minimization accuracy, with a target of $10^{-6}$ and up to two further
refinements at a tighter solver tolerance when needed. These bounds
are evaluated in floating point without rounding-error enclosures.
Every ICNN minimization meets this gap target. The initial
searches disagree by more than $10^{-6}$ for nine test-potential
calculations across seven MLP models. We repeat these searches with
$32$ initial points from each of two larger sampled sets, increasing
to $64$ for one calculation.
The additional searches confirm the retained minima, and the final
search discrepancies are below $3.3\times10^{-12}$. This agreement
does not establish global minimality for the MLP. Its reported energy errors can
therefore contain both approximation and minimization error.

\begin{figure}[htbp]
\centering
\includegraphics[width=\textwidth]{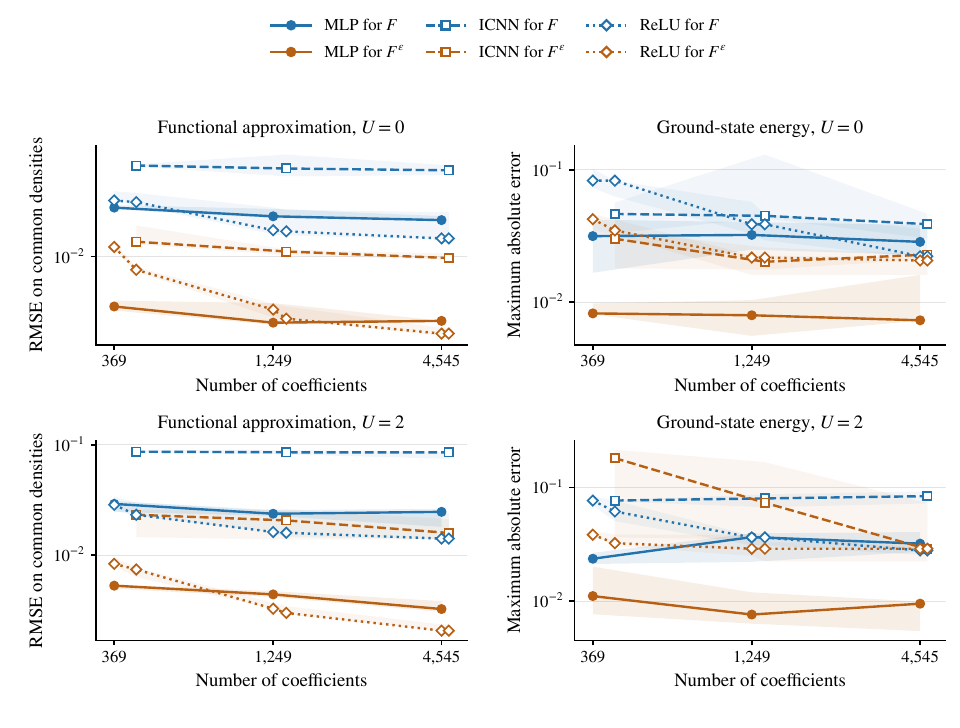}
\caption{Comparison of networks trained by MSE and Adam with the ReLU
constructions for $F$ and $F^\epsilon$, with $\epsilon=0.05$.
The left panels show the root mean-square error
on the common density test set in $\mathcal I_2$, measured against
each network's own target. The right panels show the maximum absolute
energy error over the $32$ test potentials, including the regularization
correction. Curves show the median of each per-run metric over three
seeds, and shading spans the smallest and largest values across seeds.
The horizontal axis counts trainable weights and biases for
the MLP and ICNN, and $M+1$ stored coefficients for the ReLU construction
with shared rotations. The trained widths are $16$, $32$, and $64$.
The six ReLU sizes match their coefficient counts and use a separate
pool of $2048$ sampled densities. The trained networks use $5120$ training
densities and $128$ validation densities, so matching coefficient
counts does not match the information used in fitting.}
\label{fig:num-learned}
\end{figure}

Table~\ref{tab:num-learned} and Figure~\ref{fig:num-learned} compare
the methods at comparable coefficient counts. Each group of four
rotated affine functions shares three coordinates for its zero-mean
slope and one intercept, giving $4(M/4)+1=M+1$ stored coefficients
including the constant $E(0)$. The table uses $M=1248$ for both targets,
matching the width-$32$ MLP's $1249$ parameters and remaining close
to the ICNN's $1381$. The figure varies $M$ to match both architectures
at each tested width. These larger affine constructions use a pool of
$2048$ sampled densities, fixed across $M$, retaining the original
$256$ and adding independent samples by the same rule. The greedy
selection and energy solvers are unchanged. The learned networks
use $5120$ training labels and $128$ validation labels, whereas the
affine constructions use both functional values and supporting slopes
at their $2048$ sampled densities.
The coefficient counts are therefore comparable, while the amounts
of reference information differ. All reported errors are absolute,
and each table entry and plotted curve is a median over three seeds.

\begin{table}[htbp]
\centering
\small
\caption{Median results over three random seeds for learned networks of width $32$
and ReLU constructions with $M=1248$, using $\epsilon=0.05$ and $t=1$.
Here $P$ counts trainable parameters for MLP and ICNN, and $M+1$ stored coefficients
for ReLU, shared across the four shifts of site labels.
Errors use the same $128$ densities in $\mathcal I_2$ against each network's own target
and the same $32$ potentials, including the energy correction for $F^\epsilon$.
Errors are absolute. Group minima are bold. ReLU uses $2048$ sampled densities and supporting slopes,
whereas learned networks use $5120$ training and $128$ validation labels,
so the information budgets differ.}
\label{tab:num-learned}
\begin{tabular}{@{}cllrrrrr@{}}
\hline
 & & & & \multicolumn{2}{c}{Functional error} & \multicolumn{2}{c}{Energy error}\\
$U$ & Target & Method & $P$ & RMSE & Max. & Mean & Max.\\
\hline
0 & $F$ & MLP & 1249 & 0.0224 & 0.0994 & \textbf{0.0094} & \textbf{0.0322}\\
0 & $F$ & ICNN & 1381 & 0.0580 & 0.2903 & 0.0153 & 0.0449\\
0 & $F$ & ReLU ($M=1248$) & 1249 & \textbf{0.0169} & \textbf{0.0923} & 0.0193 & 0.0388\\
\hline
0 & $F^\epsilon$ & MLP & 1249 & \textbf{0.0027} & \textbf{0.0116} & \textbf{0.0023} & \textbf{0.0080}\\
0 & $F^\epsilon$ & ICNN & 1381 & 0.0111 & 0.0303 & 0.0075 & 0.0202\\
0 & $F^\epsilon$ & ReLU ($M=1248$) & 1249 & 0.0035 & 0.0126 & 0.0094 & 0.0218\\
\hline
2 & $F$ & MLP & 1249 & 0.0237 & 0.0992 & \textbf{0.0130} & 0.0366\\
2 & $F$ & ICNN & 1381 & 0.0857 & 0.3901 & 0.0345 & 0.0801\\
2 & $F$ & ReLU ($M=1248$) & 1249 & \textbf{0.0162} & \textbf{0.0918} & 0.0175 & \textbf{0.0364}\\
\hline
2 & $F^\epsilon$ & MLP & 1249 & 0.0044 & 0.0145 & \textbf{0.0038} & \textbf{0.0076}\\
2 & $F^\epsilon$ & ICNN & 1381 & 0.0207 & 0.0685 & 0.0373 & 0.0734\\
2 & $F^\epsilon$ & ReLU ($M=1248$) & 1249 & \textbf{0.0033} & \textbf{0.0126} & 0.0110 & 0.0290\\
\hline
\end{tabular}
\end{table}

In the width-$32$ comparison in Table~\ref{tab:num-learned},
the affine constructions give the smallest functional errors for $F$
at both interaction strengths and for $F^\epsilon$ at $U=2$.
The MLP has the smallest mean energy error in all four groups
and the smallest maximum energy error except for $F$ at $U=2$.
However, we note that all three methods give
errors of comparable magnitude. Energy minimization can select
densities where the fit is less accurate than the test samples suggest,
and for $F^\epsilon$ it searches the larger set $K_{0.1}$.
The training domains differ between targets, and the activations differ
between architectures, so these comparisons do not isolate the effects
of smoothing or convexity.

In Figure~\ref{fig:num-learned}, we can see that similar to the results presented in Table \ref{tab:num-learned} the affine constructions tend to better approximate the functionals while the MLP provides better energies.

Of the $72$ selected trials, $70$ reach the
$6000$-update limit, and $46$ selected checkpoints attain their
smallest validation error at the final update. The results therefore
describe accuracy under this update cap, while the effect of longer
training remains untested.
Neither model is constrained to lie below its target, and increasing
the width does not guarantee a monotone improvement in
functional or energy accuracy.

\subsection{Summary of numerical results}
The numerical experiments illustrate the approximation framework for two spinless fermions on a four-site ring, with similar qualitative behavior in the interacting and noninteracting cases. The affine ReLU constructions preserve convexity and provide lower bounds for both functionals and energies, with errors that are nonincreasing as affine functions are added. The chosen minimization domains eliminate localization error, and the exact regularization correction allows comparison with the original ground-state energies. Accuracy nevertheless depends on the regularization parameter and the sampled density pool, and the observed plateaus show that increasing the number of affine functions alone need not substantially improve the approximation.

At comparable coefficient counts, the affine constructions generally achieve smaller sampled functional errors, while the MLP trained on the regularized functional gives the smallest energy errors. The ICNN preserves convexity but does not consistently improve energy accuracy. These results highlight that accuracy on sampled densities does not directly determine accuracy after energy minimization. The comparisons also depend on differing amounts of reference information, training domains, and a fixed training budget, and do not isolate the effects of regularization or convexity.

\section{Proofs}\label{proofsec}

\subsection{Proofs for Section~\ref{absnn}}

\begin{proof}[Proof of Theorem~\ref{sigapprox}]
Let $G\in\CF$ and let $U$ be an open neighborhood of $G$ in the compact-open
topology.  Since
$G\in\cl(\inte\CF)$, the intersection $U\cap\inte\CF$ is nonempty and open.
By Theorem~\ref{uap}, the dense set $\NF(\sigma)$ meets this intersection.  Hence
every compact-open neighborhood of $G$ contains an element of
$\NF(\sigma)\cap\inte\CF$.

For the explicit formulation, take
$U=\{H\in C(X,\RR):|H-G|_K<\delta\}$.
\end{proof}

\begin{proof}[Proof of Proposition~\ref{prop:Positivity}]
Compact-open limits preserve the inequalities defining $\CF_K^+$, so this set
is closed.  If $\min_KG>0$, the $|\cdot|_K$-ball of radius
$\frac12\min_KG$ about $G$ is contained in $\CF_K^+$.  Conversely, if
$G(\rho_0)=0$ for some $\rho_0\in K$, then $G-c$ does not belong to
$\CF_K^+$ for any $c>0$, while $G-c\to G$ in the compact-open topology as
$c\downarrow0$.
Thus $G$ is not an interior point.  Finally,
$G+n^{-1}\in\inte\CF_K^+$ and $G+n^{-1}\to G$ in the compact-open topology
for every $G\in\CF_K^+$.
\end{proof}

\begin{proof}[Proof of Theorem~\ref{thm:UAPconvex}]
Since $G$ is finite, continuous, and convex on the real normed space $X$,
the Hahn--Banach theorem gives, for each $\rho'\in K$, a subgradient
$a_{\rho'}\in\partial G(\rho')$.  The
supporting affine function
\[
 h_{\rho'}(\rho)=G(\rho')+\dua{a_{\rho'}}{\rho-\rho'}
\]
satisfies $h_{\rho'}\le G$ on $X$.  Since $G-h_{\rho'}$ is continuous and
vanishes at $\rho'$, the relatively open sets
\[
 U_{\rho'}=\{\rho\in K:G(\rho)-h_{\rho'}(\rho)<\delta\}
\]
cover $K$.  Choose a finite subcover
$U_{\rho_1},\ldots,U_{\rho_M}$ and set
\[
 H(\rho)=\max_{1\le i\le M}h_{\rho_i}(\rho).
\]
With $a_i=a_{\rho_i}$ and
$b_i=G(\rho_i)-\dua{a_i}{\rho_i}$, this is the maximum of the affine
functions appearing in \eqref{eq:convex-network}.
Then $H\le G$ on $X$ and $0\le G-H<\delta$ on $K$.  Since $G\ge0$, the
function $\NN=\max\{0,H\}$ still satisfies $\NN\le G$ on $X$ and
$0\le G-\NN<\delta$ on $K$.

The map
$\rho\mapsto(h_{\rho_1}(\rho),\ldots,h_{\rho_M}(\rho))$ is a
finite-dimensional affine feature map.  Repeated use of
\[
 \max\{s,t\}=\operatorname{ReLU}(s-t)+t
\]
computes its maximum, together with zero, by a finite-dimensional ReLU
network.  This proves the representation \eqref{eq:convex-network} and all
of the claimed structural properties.
\end{proof}

\subsection{Proofs for Section~\ref{mysec}}

\begin{proof}[Proof of Theorem~\ref{thm:MYBacho}]
The results (i)-(iii) are standard and are included for the sake of completeness.

Choose $\sigma_0\in\dom G$.  For every $\rho\in X$,
\[
 0\le G^{\epsilon}(\rho)
 \le G(\sigma_0)+\frac{1}{p\epsilon}
                    \norm{\rho-\sigma_0}_X^p.
\]
Thus the convex infimal convolution $G^{\epsilon}$ is finite and locally
bounded above. By elementary arguments~\cite[Corollary~2.2.12]{zalinescu2002convex}, convexity and local boundedness above imply local Lipschitz
continuity on smaller balls, hence continuity on $X$.
Since $G^\epsilon$ is finite, continuous, and convex on $X$,
the standard subgradient existence theorem
\cite[Theorem~2.4.9]{zalinescu2002convex} yields
$\partial G^\epsilon(\rho)\ne\varnothing$
for every $\rho\in X$.
This proves (i).

Increasing the penalty coefficient as $\epsilon$ decreases gives
\[
 G^{\epsilon}\le G^{\delta}\le G
 \qquad(0<\delta<\epsilon).
\]
Since $G(\sigma)\ge\inf_XG$ and the penalty is nonnegative, one also has
$\inf_XG\le G^{\epsilon}(\rho)$ for every $\rho\in X$.
Moreover,
\[
 \inf_{\rho\in X}G^{\epsilon}(\rho)
 =\inf_{\rho,\sigma\in X}
 \left\{G(\sigma)+\frac{1}{p\epsilon}
                       \norm{\rho-\sigma}_X^p\right\}
 =\inf_{\sigma\in X}G(\sigma),
\]
because the inner penalty is minimized by $\rho=\sigma$.  This proves (ii).

For (iii), let
$a(\sigma)=\dua{\ell}{\sigma}+\alpha\le G(\sigma)$ be a continuous affine
minorant.  Direct conjugation of the power penalty gives
\[
 G^{\epsilon}(\rho)
 \ge\inf_{\sigma\in X}
 \left\{a(\sigma)+\frac{1}{p\epsilon}\norm{\rho-\sigma}_X^p\right\}
 =a(\rho)-\frac{\epsilon^{q-1}}q\norm{\ell}_{X^*}^q.
\]
If $G$ is lower semicontinuous, the Fenchel--Moreau theorem represents
$G(\rho)$ as the supremum of its continuous affine minorants.  Taking first
$\epsilon\downarrow0$ and then the supremum yields
\[
 \liminf_{\epsilon\downarrow0}G^{\epsilon}(\rho)\ge G(\rho).
\]
Part~(ii) supplies the reverse inequality and monotonicity.

Finally, writing $z=\rho-\sigma$, the definition of infimal convolution and
\eqref{eq:power-conjugate} give, for every $v\in X^*$,
\[
 \begin{aligned}
 (G^{\epsilon})^*(v)
 &=\sup_{\sigma,z\in X}
   \left\{\dua{v}{\sigma+z}-G(\sigma)-\phi_{\epsilon}(z)\right\}\\
 &=G^*(v)+\phi_{\epsilon}^*(v)\\
 &=G^*(v)+\frac{\epsilon^{q-1}}q\norm{v}_{X^*}^q.
 \end{aligned}
\]
If the dual norm is strictly convex, the right-hand side is strictly convex.
If $v_1,v_2\in\partial G^{\epsilon}(\rho)$, the Fenchel--Young inequality
is an equality at $(v_i,\rho)$ for $i=1,2$. This forces the
conjugate to be affine on $[v_1,v_2]$, so strict convexity gives $v_1=v_2$.
Hence the nonempty subdifferential is a singleton, say
$\partial G^{\epsilon}(\rho)=\{v_\rho\}$. For a continuous convex
function $f$ on an open set, its right directional derivative satisfies
\[
 f_+'(\rho;h)=\sup_{v\in\partial f(\rho)}\dua{v}{h}.
\]
Applying this identity to both $h$ and $-h$ shows that
$G^{\epsilon}$ is G\^ateaux differentiable at $\rho$, with derivative
$v_\rho$.  This proves (iv).
\end{proof}

\begin{proof}[Proof of Proposition~\ref{prop:restricted-envelope}]
The functional $G+\iota_{\mathcal A}$ is proper, convex, and nonnegative.
Its $p$-MY regularization is the expression in
\eqref{eq:restricted-envelope}, so the claims follow from
Theorem~\ref{thm:MYBacho}.
\end{proof}

\subsection{Proofs for Section~\ref{sec:abstract-error}}

\begin{proof}[Proof of Proposition~\ref{prop:exact-affine-shift}]
Put $\eta=v|_Y\in Y^*$, which corresponds to $[v]$ under the canonical
identification $X^*/Y^\perp\simeq Y^*$.  In particular,
$\dua{\eta}{\zeta}=\dua{v}{\zeta}$ for every $\zeta\in Y$.  Since every
$\rho\in\mathcal A$ has the unique representation
$\rho=\bar\rho+\zeta$ with $\zeta\in Y$, the definition of the convex
conjugate gives
\begin{align*}
 E_{\mathcal A}(v)
 &=\inf_{\zeta\in Y}
   \left\{g(\zeta)+\dua{v}{\bar\rho}+\dua{\eta}{\zeta}\right\}\\
 &=\dua{v}{\bar\rho}
   -\sup_{\zeta\in Y}\left\{\dua{-\eta}{\zeta}-g(\zeta)\right\}\\
 &=\dua{v}{\bar\rho}-g^*(-\eta).
\end{align*}
Because $g$ is proper, $g^*(-\eta)>-\infty$. The value $+\infty$ is
possible and corresponds to $E_{\mathcal A}(v)=-\infty$.
The same calculation with $g^{\epsilon}$ in place of $g$ yields
\[
 E_{\mathcal A}^{\epsilon}(v)
 =\dua{v}{\bar\rho}-(g^{\epsilon})^*(-\eta).
\]

We next compute the conjugate of $g^{\epsilon}$.  For every
$a\in Y^*$, the change of variables $h=\zeta-\xi$ gives
\begin{align*}
 (g^{\epsilon})^*(a)
 &=\sup_{\zeta,\xi\in Y}
   \left\{\dua{a}{\zeta}-g(\xi)
   -\frac1{p\epsilon}\norm{\zeta-\xi}_Y^p\right\}\\
 &=\sup_{\xi,h\in Y}
   \left\{\dua{a}{\xi}-g(\xi)+\dua{a}{h}
   -\frac1{p\epsilon}\norm{h}_Y^p\right\}\\
 &=g^*(a)
   +\frac{\epsilon^{q-1}}q\norm{a}_{Y^*}^q,
\end{align*}
where the last step is \eqref{eq:power-conjugate} on $Y$.  Evaluating this
identity at $a=-\eta$, using
$\norm{\eta}_{Y^*}=\norm{[v]}_{Y^*}$, and substituting into
the preceding regularized-energy identity, we obtain
\begin{align*}
 E_{\mathcal A}^{\epsilon}(v)
 &=\dua{v}{\bar\rho}-g^*(-\eta)
   -\frac{\epsilon^{q-1}}q\norm{[v]}_{Y^*}^q\\
 &=E_{\mathcal A}(v)
   -\frac{\epsilon^{q-1}}q\norm{[v]}_{Y^*}^q.
\end{align*}
This proves \eqref{regeneshift} as an extended-real identity.

Finally, let $w\in Y^\perp$.  Then $\dua{w}{\zeta}=0$ for every $\zeta\in Y$.
Consequently, for either $\Phi=g$ or $\Phi=g^{\epsilon}$,
\begin{align*}
 &\inf_{\zeta\in Y}
  \left\{\Phi(\zeta)+\dua{v+w}{\bar\rho+\zeta}\right\}\\
 &\qquad=\dua{w}{\bar\rho}
  +\inf_{\zeta\in Y}
   \left\{\Phi(\zeta)+\dua{v}{\bar\rho+\zeta}\right\}.
\end{align*}
Taking $\Phi=g$ and $\Phi=g^{\epsilon}$ proves the two shift identities.
No defining infimum, in particular no proximal problem, was
assumed to be attained.
\end{proof}

\begin{proof}[Proof of Theorem~\ref{thm:ErrorLegendre}]
By Theorem~\ref{thm:MYBacho}, $g^{\epsilon}:Y\to[0,+\infty)$ is finite,
continuous, and convex.  Moreover, $K_Y=K-\bar\rho$ is compact.
Applying Theorem~\ref{thm:UAPconvex} on $Y$ with $K_Y$ gives a network
with the stated bounds.  Now let $\NN$ be any continuous network satisfying
these bounds.  The decomposition of a point in $\mathcal A=\bar\rho+Y$ is unique,
so its lift $\widetilde{\NN}$ is well defined.

Fix $v\in X^*$ with $E_{\mathcal A}(v)>-\infty$.  Since
$\mathcal A\cap\dom G\ne\varnothing$, this energy is finite.
The two objectives defining $E_K^{\epsilon}(v)$ and $E_{\NN,K}(v)$
are continuous on the compact set $K_Y$, attain their minima, and differ
uniformly by less than $\delta$.  Stability of the minimum under uniform
perturbations therefore gives
\[
 \left|E_K^{\epsilon}(v)-E_{\NN,K}(v)\right|\le\delta.
\]
The shift identity \eqref{regeneshift} makes
$E_{\mathcal A}^{\epsilon}(v)$ finite.  The triangle inequality, together
with this identity and the definition \eqref{eq:localization-error}, yields
\begin{align*}
 \left|E_{\mathcal A}(v)-E_{\NN,K}(v)
       -\frac{\epsilon^{q-1}}q\norm{[v]}_{Y^*}^q\right|
 &=\left|E_{\mathcal A}^{\epsilon}(v)-E_{\NN,K}(v)\right|\\
 &\le\left|E_{\mathcal A}^{\epsilon}(v)-E_K^{\epsilon}(v)\right|
      +\left|E_K^{\epsilon}(v)-E_{\NN,K}(v)\right|\\
 &\le\tau_{\mathcal A,K}^{\epsilon}(v)+\delta,
\end{align*}
which is \eqref{eq:absolute-energy-bound}.
\end{proof}

\subsection{Proofs for Section~\ref{sec:dft-application}}

\begin{proof}[Proof of Proposition~\ref{prop:dft-local-blowup}]
Put $u=\sqrt\rho$.  Since $\int_{\RR^3}\rho=N>0$, choose
$\eta\in C_c^\infty(\RR^3)$ with $0\le\eta\le1$ and
\begin{equation*}
 m_1=\int_{\RR^3}\eta\rho>0,
 \qquad
 m_2=\int_{\RR^3}\eta^2\rho>0.
\end{equation*}
For example, the second inequality follows from
$m_1^2\le N m_2$.  For $j\ge1$, set
\begin{equation*}
 s_j(x)=\sin(j^2x_1),
 \qquad
 \alpha_j=\frac1{m_1}\int_{\RR^3}\eta\rho s_j,
 \qquad
 q_j=\frac1{4j}\eta(s_j-\alpha_j).
\end{equation*}
Since $|s_j|\le1$ and $\eta\rho\ge0$,
\begin{equation*}
 |\alpha_j|\le1,
 \qquad
 |q_j|\le\frac1{2j},
 \qquad
 \int_{\RR^3}q_j\rho=0.
\end{equation*}
Define
\begin{equation*}
 \chi_j=(1+q_j)^{1/2},
 \qquad
 \rho_j=\chi_j^2\rho=(1+q_j)\rho.
\end{equation*}
Then
\begin{equation*}
 \left(1-\frac1{2j}\right)\rho
 \le\rho_j
 \le\left(1+\frac1{2j}\right)\rho,
 \qquad
 \int_{\RR^3}\rho_j=N.
\end{equation*}
Moreover,
\begin{equation}\label{eq:dft-blowup-chi-bounds}
 \frac1{\sqrt2}\le\chi_j\le\sqrt{\frac32},
 \qquad
 \sqrt{\frac23}\le\chi_j^{-1}\le\sqrt2.
\end{equation}
Because $\chi_j-1\in C_c^\infty(\RR^3)$ and
$\chi_j\in W^{1,\infty}(\RR^3)$, the multiplier property of
$W^{1,\infty}$ on $H^1$ gives
\begin{equation*}
 \sqrt{\rho_j}=\chi_j u\in H^1(\RR^3).
\end{equation*}
Thus $\rho_j\in\mathcal I_N$.  The pointwise estimate
\begin{equation*}
 |\rho_j-\rho|=|q_j|\rho\le\frac1{2j}\rho
\end{equation*}
gives both asserted norm bounds.

It remains to prove the kinetic-energy divergence.  Write
\begin{equation*}
 P_j=u\nabla\eta(s_j-\alpha_j),
 \qquad
 Q_j=j^2\eta u\cos(j^2x_1)e_1.
\end{equation*}
Differentiating $\chi_j$ gives
\begin{equation}\label{eq:dft-blowup-gradient}
 u\nabla\chi_j
 =\frac1{8j}\chi_j^{-1}(P_j+Q_j),
 \qquad
 \norm{P_j}_{L^2}\le2\norm{u\nabla\eta}_{L^2}.
\end{equation}
On the other hand,
\begin{align*}
 \norm{\eta u\cos(j^2x_1)}_{L^2}^2
 &=\frac{m_2}{2}
   +\frac12\int_{\RR^3}\eta^2\rho\cos(2j^2x_1)\\
 &\longrightarrow\frac{m_2}{2}
\end{align*}
by the Riemann--Lebesgue lemma.  Hence, for all sufficiently large $j$,
\begin{equation}\label{eq:dft-blowup-high-frequency}
 \norm{Q_j}_{L^2}
 =j^2\norm{\eta u\cos(j^2x_1)}_{L^2}
 \ge\frac{j^2\sqrt{m_2}}2.
\end{equation}
Using \eqref{eq:dft-blowup-chi-bounds}--\eqref{eq:dft-blowup-high-frequency}
and the reverse triangle inequality, we obtain
\begin{align*}
 \norm{u\nabla\chi_j}_{L^2}
 &\ge\frac1{8j}
 \left(
   \sqrt{\frac23}\norm{Q_j}_{L^2}
   -\sqrt2\norm{P_j}_{L^2}
 \right)\\
 &\longrightarrow+\infty.
\end{align*}
At the same time,
\begin{equation*}
 \norm{\chi_j\nabla u}_{L^2}
 \le\sqrt{\frac32}\norm{\nabla u}_{L^2}.
\end{equation*}
The weak product rule and another application of the reverse triangle
inequality therefore yield
\begin{equation*}
 \norm{\nabla\sqrt{\rho_j}}_{L^2}
 =\norm{u\nabla\chi_j+\chi_j\nabla u}_{L^2}
 \longrightarrow+\infty.
\end{equation*}
Finally, every canonical state can be regarded as a grand-canonical state
supported in the $N$-particle sector, so
$F_{\mathrm L}(\rho_j)\ge F_{\mathrm{GC}}(\rho_j)$.  The second inequality
and the claimed divergence follow from
\eqref{eq:dft-ho-grand-canonical}.
\end{proof}

\begin{proof}[Proof of Corollary~\ref{cor:dft-energy-approximation}]
Apply Theorem~\ref{thm:ErrorLegendre} with
\[
 G=F_{\mathrm L},\qquad
 \mathcal A=\mathcal A_N=\bar\rho+Y,
 \qquad
 K\subset\mathcal I_N.
\]
The space $X$ is separable, $F_{\mathrm L}$ is proper, nonnegative, and
convex, and $\bar\rho\in\mathcal A_N\cap\dom F_{\mathrm L}$.  Moreover,
$\dom F_{\mathrm L}\subset\mathcal A_N$, so
\[
 E_{\mathcal A_N}(v)=E(v),
 \qquad
 E_{\mathcal A_N}^{\epsilon}(v)=E^{\epsilon}(v).
\]
Finally, \eqref{eq:dft-annihilator} and \eqref{eq:dft-quotient-norm}
identify the regularization correction in Theorem~\ref{thm:ErrorLegendre} with
$\phi_{\epsilon,Y}^*([v])$.  Under these identifications,
\eqref{eq:absolute-energy-bound} is exactly
\eqref{eq:dft-absolute-energy-estimate}.
\end{proof}

\begin{proof}[Proof of Proposition~\ref{prop:dft-network-energy}]
For each $\rho=\bar\rho+\zeta\in K$, the map
\[
 v\longmapsto \NN(\zeta)+\dua{v}{\rho}
\]
is finite and affine.  Its pointwise minimum over the compact set
$K$ is finite and attained because the objective is continuous in $\rho$. As
an infimum of affine functions of $v$, it is concave.  For $v,w\in X^*$, the elementary inequality
between two minima gives
\begin{align*}
 |E_{\NN,K}(v)-E_{\NN,K}(w)|
 &\le\max_{\rho\in K}|\dua{v-w}{\rho}|\le\left(\max_{\rho\in K}\norm{\rho}_X\right)
       \norm{v-w}_{X^*},
\end{align*}
which proves \eqref{eq:dft-network-energy-lipschitz}.  Since every
$\rho\in K$ has mass $N$, adding $c\,1$ to the potential adds $cN$ to
every objective in \eqref{eq:dft-network-energy}. This proves
\eqref{eq:dft-network-energy-shift}.  Finally, $K\subset\mathcal I_N$
consists of nonnegative densities.  Thus $v\le w$ implies
$\dua{v}{\rho}\le\dua{w}{\rho}$ for every $\rho\in K$, and taking minima
proves order preservation.
\end{proof}

\begin{proof}[Proof of Corollary~\ref{cor:dft-gc-energy-approximation}]
Apply Theorem~\ref{thm:ErrorLegendre} with
\begin{equation*}
 G=F_{\mathrm{GC}},
 \qquad
 \mathcal A=\mathcal A_\nu=\bar\rho_\nu+Y.
\end{equation*}
The space $X$ is separable, $F_{\mathrm{GC}}$ is proper, nonnegative, and
convex, and
$\bar\rho_\nu\in\mathcal A_\nu\cap\dom F_{\mathrm{GC}}$.  The definitions
give
\begin{equation*}
 E_{\mathcal A_\nu}(v)
 =E_\nu^{\mathrm{GC}}(v),
 \qquad
 E_{\mathcal A_\nu}^{\epsilon}(v)
 =E_\nu^{\mathrm{GC},\epsilon}(v).
\end{equation*}

For completeness, the exact energy is finite for every $v\in X^*$.  Indeed,
the infinitesimal one-body form bound gives, for every $a>0$,
\begin{equation*}
 |\dua{v}{\rho}|
 \le a\norm{\nabla\sqrt\rho}_{L^2}^2+C_{a,v}\nu,
 \qquad \rho\in\mathcal I_\nu.
\end{equation*}
Taking $a<1$ and using \eqref{eq:dft-ho-grand-canonical} yields a finite
lower bound for $E_\nu^{\mathrm{GC}}(v)$, while the competitor
$\bar\rho_\nu$ yields a finite upper bound.  Finally,
\eqref{eq:dft-annihilator} and \eqref{eq:dft-quotient-norm} identify the
regularization correction in Theorem~\ref{thm:ErrorLegendre} with
$\phi_{\epsilon,Y}^*([v])$.  Under these identifications,
\eqref{eq:absolute-energy-bound} is exactly
\eqref{eq:dft-gc-absolute-energy-estimate}.
\end{proof}

\begin{proof}[Proof of \eqref{eq:dft-zero-potential-localization}]
Dilating a compactly supported, finite-energy $N$-particle trial state
makes its kinetic and Coulomb energies tend to zero.  Nonnegativity
therefore gives $E(0)=0$, and \eqref{eq:dft-regularized-shift} gives
$E^{\epsilon}(0)=0$ as well.
The Hoffmann--Ostenhof bound implies $F_{\mathrm L}(\rho)>0$ for every
$\rho\in\mathcal I_N$: otherwise $\sqrt\rho$ would be a constant in
$L^2(\RR^3)$ with positive mass.  Lower semicontinuity and compactness
then give $\min_KF_{\mathrm L}>0$.
If $F_{\mathrm L}^{\epsilon}(\rho)=0$, a sequence in its defining
infimum would converge to $\rho$ in $X$ with universal energies tending
to zero, contradicting lower semicontinuity.  Continuity of the envelope
thus gives $\min_KF_{\mathrm L}^{\epsilon}>0$.

To prove convergence of the minima, take $\epsilon_j\downarrow0$ and
minimizers $\rho_j\in K$ of $F_{\mathrm L}^{\epsilon_j}$ on $K$.
After passing to a subsequence, $\rho_j\to\rho_*\in K$.
For every fixed $a>0$, eventually $\epsilon_j<a$, so monotonicity and
continuity give
\[
 \lim_{j\to\infty}\min_KF_{\mathrm L}^{\epsilon_j}
 \ge\lim_{j\to\infty}F_{\mathrm L}^{a}(\rho_j)
 =F_{\mathrm L}^{a}(\rho_*).
\]
Letting $a\downarrow0$ yields a lower bound of
$F_{\mathrm L}(\rho_*)\ge\min_KF_{\mathrm L}$. The reverse bound follows
from $F_{\mathrm L}^{\epsilon}\le F_{\mathrm L}$.
Together with the definition of $\tau_K^{\epsilon}(0)$, this proves
\eqref{eq:dft-zero-potential-localization}.

For fixed mean mass $\nu>0$, mix a dilated $k$-particle trial state,
$k\ge\nu$, with the vacuum state ($\Gamma_0=1$ and $\Gamma_n=0$ for $n\ge1$)
using weights $\nu/k$ and $1-\nu/k$.
Its internal energy tends to zero, proving $E_\nu^{\mathrm{GC}}(0)=0$.
The same argument applies to $F_{\mathrm{GC}}+\iota_{\mathcal A_\nu}$,
using \eqref{eq:dft-ho-grand-canonical} and
\eqref{eq:dft-gc-regularized-shift}, and proves the stated grand-canonical
extension.  Finally, the uniform approximation in
\eqref{eq:dft-network-approximation} bounds the difference between
$E_{\NN,K}(0)$ and $\min_KF_{\mathrm L}^{\epsilon}$ by $\delta$.
Thus $E_{\NN,K}(0)\to\min_KF_{\mathrm L}>0$ for fixed $K$ as
$\epsilon,\delta\to0$.
\end{proof}

\begin{proof}[Proof of Proposition~\ref{prop:dft-compact-potential-classes}]
For each $v\in\mathcal V$, choose $\rho_v\in\mathcal I_N$ such that
\[
 F_{\mathrm L}(\rho_v)+\dua{v}{\rho_v}<E(v)+\eta/2.
\]
Continuity of $E$ and of the pairing with $\rho_v$ gives a neighborhood
$U_v$ of $v$ on which
$F_{\mathrm L}(\rho_v)+\dua{w}{\rho_v}<E(w)+\eta$.
Choose a finite subcover $U_{v_1},\ldots,U_{v_J}$ of $\mathcal V$ and set
$K_\eta=\operatorname{conv}\{\rho_{v_1},\ldots,\rho_{v_J}\}$.
This set is compact and convex, is independent of $\epsilon$ and $\delta$,
and lies in the convex effective domain $\mathcal I_N$ of $F_{\mathrm L}$.
Lower semicontinuity gives attainment of the exact energy minimum on
$K_\eta$, and the selected competitors prove
\eqref{eq:dft-compact-class-localization}.

For the network from Corollary~\ref{cor:dft-energy-approximation},
underestimation of $E(v)$ is at most $\phi_{\epsilon,Y}^*([v])+\delta$:
the uniform approximation changes the minimum on $K_\eta$ by at most
$\delta$, and the regularized minimum on $K_\eta$ is bounded below by
the unrestricted value $E(v)-\phi_{\epsilon,Y}^*([v])$.
Overestimation is at most $\eta$, since
$\NN(\rho-\bar\rho)\le F_{\mathrm L}^{\epsilon}(\rho)
\le F_{\mathrm L}(\rho)$ on $K_\eta$ and
\eqref{eq:dft-compact-class-localization} bounds the exact localization error.
Taking the maximum of these two error magnitudes and using
\[
 \sup_{v\in\mathcal V}\phi_{\epsilon,Y}^*([v])
 =\frac{\epsilon^{q-1}}q\max_{v\in\mathcal V}\norm{[v]}_{Y^*}^q
\]
proves
\eqref{eq:dft-compact-class-energy-bound} and the uniform convergence
claim.  The construction uses only approximate minimizing densities.

For the grand-canonical case, $-E_\nu^{\mathrm{GC}}$ is a finite
lower-semicontinuous convex functional on the Banach space $X^*$ and is
therefore continuous.  Replacing $F_{\mathrm L}$, $E$, and $\mathcal I_N$
by $F_{\mathrm{GC}}$, $E_\nu^{\mathrm{GC}}$, and $\mathcal I_\nu$,
respectively, the same argument applies with
Corollary~\ref{cor:dft-gc-energy-approximation} and the fixed-mass shift
\eqref{eq:dft-gc-regularized-shift}.
\end{proof}

\bibliographystyle{plain}
\bibliography{dft}

\end{document}